\documentclass[journal]{IEEEtran}
\IEEEoverridecommandlockouts
\usepackage{cite}
\usepackage{amsmath,amssymb,amsfonts}
\usepackage{algorithmic}
\usepackage{amsmath} 
\usepackage{graphicx}
\usepackage{textcomp}
\usepackage{xcolor}
\usepackage{amsmath}
\renewcommand{\vec}[1]{\boldsymbol{#1}}
\usepackage{algorithm}
\usepackage{algorithmic}

\usepackage{graphicx} 
\usepackage{float} 
\usepackage{stfloats}
\usepackage[section]{placeins}
\usepackage{subfigure}
\usepackage{booktabs}
\usepackage[numbers,sort&compress]{natbib}
\usepackage{amssymb}
\usepackage{setspace}
\usepackage{multirow}
\usepackage{color}
\usepackage{pgfplots}
\fontsize{5.0pt}{\baselineskip}\selectfont
\usepackage{epsfig}
\def\BibTeX{{\rm B\kern-.05em{\sc i\kern-.025em b}\kern-.08em
    T\kern-.1667em\lower.7ex\hbox{E}\kern-.125emX}}

\begin{document}
\pgfplotsset{compat=1.14}
\addtolength{\topmargin}{0.04in}
\title{ Weighted SPICE Algorithms  for  Range-Doppler Imaging Using One-Bit  Automotive Radar  
}

\author{Xiaolei~Shang, ~Jian~Li,~\IEEEmembership{Fellow,~IEEE},  and ~Petre~ Stoica,~\IEEEmembership{Fellow,~IEEE}
\thanks{This work was supported in part by the  National Natural Science Foundation of China under Grant 61771442, in part by Key Research Program of Frontier Sciences of CAS under Grant QYZDY-SSW-JSC035, and in part by the Swedish Research Council (VR grants 2017-04610  and 2016-06079).}
\thanks{X. Shang is  with the Department of Electronic Engineering and Information
Science, University of Science and Technology of China, Hefei 230027, China
(e-mail: xlshang@mail.ustc.edu.cn).}
\thanks{J. Li is with the Department of Electrical and Computer Engineering, University
of Florida, Gainesville, FL 32611 USA (e-mail: li@dsp.ufl.edu). }
\thanks{P. Stoica is with the Department of Information Technology, Uppsala University, Uppsala SE-751 05, Sweden (e-mail: ps@it.uu.se).} }
\maketitle
\begin{abstract}
We consider the problem of range-Doppler imaging using one-bit automotive  LFMCW\footnote{All abbreviations used in this paper are explained at the end of the Introduction.} or PMCW radar that  utilizes  one-bit ADC sampling with time-varying thresholds at the  receiver. The one-bit sampling technique can significantly reduce the cost as well as the power consumption of automotive radar systems. We formulate the one-bit LFMCW/PMCW  radar range-Doppler imaging problem as  one-bit sparse parameter estimation. The recently proposed  hyperparameter-free  (and hence user friendly) weighted SPICE algorithms, including SPICE, LIKES, SLIM and IAA, achieve excellent parameter estimation performance for data sampled with  high precision. However, these  algorithms  cannot  be  used directly for one-bit data. In this paper we first present a regularized minimization algorithm, referred  to as 1bSLIM, for accurate range-Doppler imaging using one-bit radar systems. Then, we describe how  to extend the SPICE, LIKES and IAA algorithms to the one-bit data case,  and refer to these extensions as 1bSPICE, 1bLIKES and 1bIAA. These one-bit hyperparameter-free algorithms are unified within  the one-bit weighted SPICE framework.  Moreover, efficient  implementations of the aforementioned algorithms are investigated  that rely heavily on the use of FFTs. Finally, both simulated and experimental  examples are provided to demonstrate the effectiveness of the proposed algorithms  for  range-Doppler imaging using one-bit automotive radar systems. 
\end{abstract}
\section{Introduction}
\IEEEPARstart{M}{illimeter} wave radar is   widely used in diverse applications including  advanced driver assistance automotive  systems and  fully autonomous vehicles \cite{meinl2017experimental}. Compared with other sensing systems, such as cameras, lidars and ultrasonics, radar systems can provide better performance, especially in poor lighting or adverse weather conditions  \cite{bilik2019rise}. \\
\indent The most commonly used  probing waveform in existing commercial automotive radar systems  is the LFMCW waveform,  which reduces the system cost by allowing low-rate ADCs at the radar receivers. An LFMCW radar system transmits a series of chirps, which are reflected by the targets and mixed with the transmitted chirp at the receiver. The range or range-Doppler estimation problem is converted to a 1-D or 2-D complex sinusoidal parameter estimation problem \cite{patole2017automotive,engels2017advances}. However, mutual interferences among LFMCW automotive radars become more severe as these radar systems become more widely used, since they  occupy the same 77-81 GHz band.  PMCW waveforms can be used to mitigate the interference problems. To reduce costs, binary sequences with desirable auto- and cross-correlation properties have been designed for PMCW automotive radars (see, e.g., \cite{lin2019efficient} and the references therein).  The  flexibility and diversity offered by  binary sequences allow the PMCW radars to mitigate mutual interferences  similar to the CDMA communication systems \cite{alland2019interference}. Also, near orthogonality  can be realized in the coding domain, which is advantageous for  MIMO operations \cite{li2009mimo,ertan2020spatial}.  \\
\indent To improve  the range  resolution  of automotive radars, the fundamental approach is to increase the bandwidth of the transmitted signal, which  can increase the ADC cost and power consumption.  For example, a PMCW radar usually performs demodulation in the digital domain \cite{hakobyan2019high}. A  PMCW radar with a chip rate of 2 GHz requires a sampling rate of at least 2 GHz  \cite{giannini201479}.  However,  an  ADC  with  high  precision quantization and high sampling rate is both expensive and power hungry. Yet low cost and low power consumption are crucial for commercial radar systems such as automotive radar.  \\
\indent One-bit sampling is a promising technique to mitigate the aforementioned ADC problems because of its low cost and low power consumption advantages.  Due to these attractive properties, one-bit sampling  has  been considered for radar sensing \cite{zhang2019range,zhao2019one}, spectral estimation \cite{li2014robust,host2000effects,heng}, as well as massive MIMO communications \cite{li2017channel,choi2016near,mo2017channel}. The conventional one-bit sampling considered in, e.g.,  \cite{bar2002doa,yan2012robust,knudson2016one}, compares the original signal with a zero threshold, which results in a loss of signal amplitude information. Recently, time-varying thresholds have been considered for one-bit sampling to achieve  accurate amplitude estimation \cite{ren2019sinusoidal,heng}. In such a case, signed measurements are obtained by comparing the received signal with a known time-varying threshold.  \\
\indent In this paper, we consider one-bit sampling  with time-varying thresholds for  automotive radars to reduce  cost  and power consumption. We formulate the one-bit range-Doppler estimation problem as a one-bit sparse parameter estimation problem. Sparse parameter estimation using  high-precision data has attracted  much  attention in the last two decades. Many  algorithms have been developed with the goal of obtaining accurate sparse parameter estimates. Among them, the unified weighted SPICE algorithms \cite{stoica2014weighted}, including SPICE \cite{stoica2010new}, LIKES \cite{stoica2012spice}, SLIM \cite{tan2010sparse} and IAA  \cite{yardibi2010source,roberts2010iterative,xue2011iaa}, do not require the tuning of  any user-parameter and hence are easy to use in practice. These algorithms yield excellent estimation performance, in particular they possess high resolution and low sidelobe properties. Furthermore,  IAA works well in various environments and the robustness of IAA has been ranked  high in a recent review article \cite{sun2020mimo}. However, these algorithms, together with the conventional FFT,  cannot be used directly  to  solve one-bit sparse parameter estimation problems. Herein, we first present a regularized minimization algorithm, referred  to as 1bSLIM,  for accurate range-Doppler imaging using one-bit automotive radar. Additionally, we show that 1bSLIM, devised by means of a majorization-minimization  approach \cite{hong2015unified,sun2016majorization,stoica2004cyclic,hunter2004tutorial,mairal2015incremental}, can be interpreted as applying the conventional SLIM algorithm to certain modified high-precision data. Inspired by this relationship, we describe how to extend  SPICE, LIKES and IAA to the one-bit case, and refer to these extensions as 1bSPICE, 1bLIKES and 1bIAA. Each of the algorithms we present herein has its own merits and may be preferred in a specific application.  Similar to the original  weighted SPICE algorithms, which unify SPICE, LIKES, SLIM and IAA under the same weighted SPICE umbrella,  the aforementioned one-bit algorithms are unified within the one-bit weighted SPICE framework.  Moreover, these four one-bit algorithms are all hyperparameter-free and thus are easy to use in practice.  Finally, simulated and experimental examples are provided to demonstrate the effectiveness of the proposed algorithms for range-Doppler imaging  using one-bit automotive radars. \\
\indent \textit{Notation:} We denote vectors and matrices by bold lowercase and uppercase letters, respectively. $(\cdot)^T$ and $(\cdot)^H$ represent the transpose and the conjugate transpose, respectively. $\vec{R}\in \mathbb{R}^{N\times M}$ or  $\vec{R}\in \mathbb{C}^{N\times M}$ denotes a real or complex-valued $N\times M$ matrix $\vec{R}$. $\otimes$ denotes   the Kronecker matrix product. ${\rm vec}(\cdot)$ refers to the column-wise  vectorization operation and ${\rm diag}(\vec{d})$ denotes a diagonal matrix with diagonal entries formed  from $\vec{d}$. $\Vert \cdot \Vert$ and $\Vert \cdot \Vert_F$ are, respectively, the $\ell_2$ and Frobenius norms of vectors and matrices. ${\rm tr}\lbrace\vec{R}\rbrace$ and $\left|\vec{R}\right|$, respectively, denote the trace and determinant of a square matrix $\vec{R}$. $\vec{A} \preceq \vec{B}$ means that  $\vec{B}-\vec{A}$ is a positive semidefinite matrix.   ${\rm Re}[x]$ and ${\rm Im}[x]$ denote the real- and imaginary-parts of $x$. $x$ mod $N$ is the
remainder of the Euclidean division of $x$ by $N$.   Finally, $j=\sqrt{-1}$. \\
\indent Throughout this paper, we use the following abbreviations:
\vspace{-0.5cm}
\begin{table}[h]
\footnotesize
\centering
\linespread{1.5} 
\scalebox{0.92}{
\begin{tabular}{l|l}
\toprule
ADC & Analog-to-digital converter \\
AGC & Automatic gain control\\
CDMA & Code division multiple access \\
CGLS & Conjugate-gradient least squares \\
CPI & Coherent  processing  interval \\
DAC & Digital-to-analog converter \\
DFT & Discrete Fourier transform \\
FFT  & Fast Fourier transform \\
IAA & Iterative adaptive approach \\
LIKES &  Likelihood-based estimation of sparse parameters \\
LFMCW & Linear frequency-modulated continuous-wave \\
LMMSE & Linear  minimum mean square error \\
MAP & Maximum \textit{ a posteriori} \\
MIMO & Multiple-input multiple-output \\
MM & Majorization-minimization \\
NMSE & Normalized mean-squared error \\
PMCW & Phase-modulated continuous-wave \\
PRI & Pulse repetition interval \\
SLIM & Sparse learning via iterative minimization \\
SPICE & Sparse iterative covariance-based estimation \\
SNR  & Signal-to-noise ratio \\
\bottomrule
\end{tabular} }
\end{table}
\section{Problem Formulation}
In this section,   we first briefly describe  the  signal models for the LFMCW and PMCW radars. Then, using  the signed measurements  obtained by one-bit sampling with time-varying thresholds, we formulate the range-Doppler imaging problem as a  sparse parameter estimation problem. 
\subsection{LFMCW Radar} \label{section:LFMCW}
Without loss of generality, we consider an LFMCW single-input single-output  radar. A simplified model for the received signal (after dechirping and pre-processing) is as follows (see e.g.,  \cite{zhang2019range}):
\begin{equation}
    y(n_1,n_2)=\sum_{k_{\rm d}=1}^{K_{\rm d}} \sum_{k_{\rm r}=1}^{K_{\rm r}} \gamma_{k_{\rm r},k_{\rm d}} e^{j(\omega_{k_{\rm r}}n_1+\bar{\omega}_{k_{\rm d}}n_2)}  + e(n_1,n_2),\label{eq:LFMCW}
\end{equation}
where $K_{\rm r}$ and $K_{\rm d}$ are the numbers of the grid points in the range and Doppler domains, respectively; the fast-time index $n_1, n_1=1,2,\dots, N_1$, corresponds to the $n_1$th sample within a chirp; the slow-time index $n_2, n_2=1,2,\dots, N_2$, is for the $n_2$th chirp; $\gamma_{k_{\rm r},k_{\rm d}}$  represents the reflection coefficient of a scatterer corresponding to the frequency pair $(\omega_{k_{\rm r}},\bar{\omega}_{k_{\rm d}})$; and $e(n_1,n_2)$ is the unknown additive noise. The frequency pair $(\omega_{k_{\rm r}},\bar{\omega}_{k_{\rm d}})$ is approximately related to the  range-Doppler pair $(R_{k_{\rm r}},v_{k_{\rm d}})$ through the following equations:
\begin{equation}
    \omega_{k_{\rm r}}=2\pi\left(\frac{4\mu R_{k_{\rm r}}}{cf_{\rm s}}+\frac{2f_0v_{k_{\rm d}}}{cf_{\rm s}}\right), \quad \bar{\omega}_{k_{\rm d}}=\frac{4\pi f_0v_{k_{\rm d}}T_{\rm c}}{c}, \label{eq:relation}
\end{equation}
where $2\mu$ is the chirp rate; $c$ is the speed of light; $f_{\rm s}$ is the sampling frequency; $f_0$ denotes the carrier frequency; and  $T_{\rm c}$ is the chirp duration. The range-Doppler information can therefore be recovered from the frequency pair $(\omega_{k_{\rm r}},\bar{\omega}_{k_{\rm d}})$ by using Equation (\ref{eq:relation}). The range-Doppler estimation problem associated with  (\ref{eq:LFMCW}) for an LFMCW radar is equivalent to  a 2-D  complex-valued sinusoidal parameter estimation problem.    Let the measurement vector be denoted  $\vec{y}=\begin{bmatrix} y(1,1),  y(2,1),  \dots,  y(N_1,N_2)\end{bmatrix}^T$ and let $\vec{e}$ denote the unknown noise vector. Let the unknown parameter vector be denoted  $\vec{\gamma}=\begin{bmatrix} \gamma_{1,1}, \gamma_{2,1}, \dots, \gamma_{K_{\rm r},K_{\rm d}}\end{bmatrix}^T$. Equation (\ref{eq:LFMCW}) can then be reformulated as:
\begin{equation}
    \vec{y}=\vec{B}\vec{\gamma}+\vec{e} \in \mathbb{C}^{N}, \quad \vec{B} \in  \mathbb{C}^{N\times M}, \quad M \gg N, \label{eq:linear}
\end{equation}
where $N=N_1N_2$ and $M=K_{\rm r}K_{\rm d}$. $K_{\rm r}\gg N_1$ and $K_{\rm d} \gg N_2$ are  typical assumptions used by sparse parameter estimation algorithms to form  a sufficiently fine grid,  which enables us to obtain high resolution spectral estimates and  excellent results even when the true target frequencies do not lie on the grid. Let $\vec{\psi}_{N}(\omega)=\begin{bmatrix} 1, \  e^{j\omega}, \dots, e^{j(N-1)\omega} \end{bmatrix}^T$. Then, the columns of $\vec{B}$ are given by $\lbrace \vec{\psi}_{N_1,N_2}\left( \omega_{k_{\rm r}},\bar{\omega}_{k_{\rm d}} \right) \rbrace_{k_{\rm r}=1,k_{\rm d}=1}^{K_{\rm r},K_{\rm d}}$ with $\vec{\psi}_{N_1,N_2}(\omega_{k_{\rm r}},\bar{\omega}_{k_{\rm d}})=\vec{\psi}_{N_2}(\bar{\omega}_{k_{\rm d}})\otimes \vec{\psi}_{N_1}(\omega_{k_{\rm r}})$.
\par For radar range-Doppler imaging applications, the number of potential scatterers  of interest in a scene is usually much smaller than the number of grid points, which means that most of the elements in $\vec{\gamma}$ will be zero. Hence, the LFMCW radar range-Doppler imaging problem in (\ref{eq:linear})  can be treated  as a sparse parameter estimation problem.
\begin{figure}[htb]
    \centering
    \includegraphics[width=0.48\textwidth]{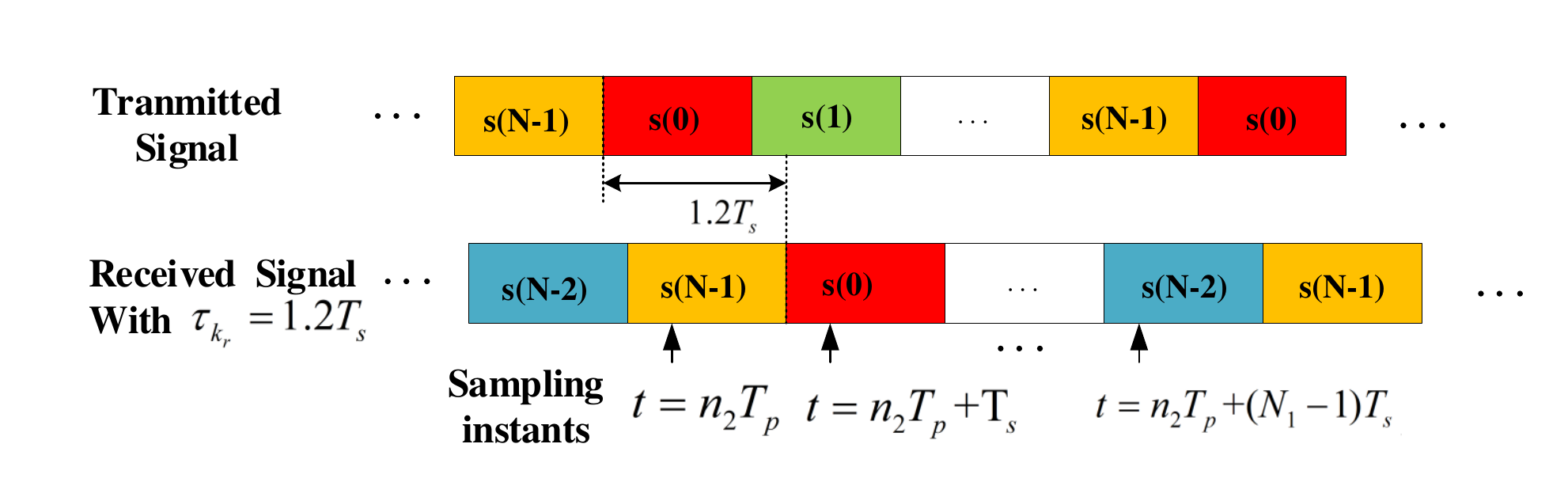}
    \caption{The transmitted sequence and the received sequence with $\tau_{k_r}=1.2T_s$.}
    \label{fig:transmit}
\end{figure} 
\subsection{PMCW Radar}
The  PMCW radar periodically transmits  a binary sequence $\vec{s}\!=\begin{bmatrix} s(0), \ s(1),  \dots,  s(N_1-1)\end{bmatrix}^T$, where $N_1$ is the sequence length and each element of $\vec{s}$ is either $1$ or $-1$, i.e., $s(n_1)\in \lbrace 1,-1\rbrace$.  Then, the transmitted baseband signal can be written as:
\begin{equation}
x(t)=\sum_{n_2=-\infty}^{\infty}\sum_{n_1=0}^{N_1-1} s(n_1)p(t-n_1T_{\rm s}-n_2T_{\rm p}),
\end{equation}
where $T_{\rm s}$ is the so-called chip duration; $T_{\rm p}=N_1T_{\rm s}$ is the PRI; and  $p(t)$ is the impulse response of a pulse-shaping filter which, without loss of generality, is chosen as a raised-cosine with a roll-off factor of 0.25. For the convenience of the subsequent signal processing, the pulse shaping filter is assumed to be non-zero in the interval $\left(-5T_{\rm s},5T_{\rm s}\right]$. Let the range and Doppler domains within the CPI be grided into $K_{\rm r}$ and $K_{\rm d}$ grid points, respectively. Then, at the receiver, we have: 
\begin{equation}
y(t)=\sum_{k_{\rm r}=1}^{K_{\rm r}}\sum_{k_{\rm d}=1}^{K_{\rm d}}\gamma_{k_{\rm r},k_{\rm d}}x(t-\tau_{k_{\rm r}})e^{j2\pi f_{k_{\rm d}}t}+e(t),
\end{equation}
where $\tau_{k_{\rm r}}$, $f_{k_{\rm d}}$, and $\gamma_{k_{\rm r},k_{\rm d}}$ represent the round-trip time delay, the Doppler frequency, and  the corresponding reflection coefficient, respectively; and $e(t)$ is the unknown additive noise. For a real-valued time delay $\tau_{k_r}$, we can split it into the integer part and the fractional part, i.e., $\tau_{k_{\rm r}} =l_{k_{\rm r}}T_{\rm s} +\bar{\tau}_{k_{\rm r}}$, where $l_{k_{\rm r}}=\lfloor \tau_{k_{\rm r}}/T_{\rm s} \rfloor$ with $\lfloor \cdot \rfloor$ being the floor operation.  Note that the echo sequence with time delay $\tau_{k_{\rm r}} =l_{k_{\rm r}}T_{\rm s} +\bar{\tau}_{k_{\rm r}}$  circularly shifts the transmitted sequence by $l_{k_r}$  positions, as shown in Fig. \ref{fig:transmit}. The echo signal with a real-valued time delay  $\tau_{k_{\rm r}}$ for the $n_2$th period (i.e., $n_2T_{\rm p}\leq t < (n_2+1)T_{\rm p}$) can be written as:
\begin{align}
    x(t-\tau_{k_{\rm r}})= &\sum_{i=-5}^{N_1+4}s\left(({N_1}+i-l_{k_{\rm r}})\ {\rm mod} \ N_1\right) \nonumber\\
    & \qquad   \cdot p\left(t-n_2T_{\rm p} - i T_{\rm s}-\bar{\tau}_{k_{\rm r}}\right). \label{eq:appro}
\end{align}
We assume that  there are $N_2$ periods during one CPI and the Doppler shift within one period is so small  that it can be neglected \cite{giannini201479,hakobyan2019high}. Then, the received baseband signal for the $n_2$th period  can be written as:
\begin{align}
y_{n_2}(t)=\sum_{k_{\rm r},k_{\rm d}}\sum_{i=-5}^{N_1+4}\gamma_{k_{\rm r},k_{\rm d}}s\left(({N_1}+i-l_{k_r})\ {\rm mod} \ N_1\right) \nonumber \\
\cdot p\left(t-n_2T_{\rm p} -i T_{\rm s}-\bar{\tau}_{k_r}\right)e^{j\bar{\omega}_{k_{\rm d}}n_2}+e(t),
\end{align}
where $\bar{\omega}_{k_{\rm d}}=2\pi f_{k_{\rm d}}T_{\rm p}$. We insert $t=n_2T_{\rm p}+n_1T_{\rm s}$ into $x(t-\tau_{k_{\rm r}})$ and define:
\begin{small}
\begin{align}
    \bar{s}(n_1,\tau_{k_r}) &\triangleq x(n_2T_p+n_1T_s-\tau_{k_r}) \nonumber \\
    &= \sum_{i=-5}^{4}s\left( (N_1+n_1+i-l_{k_{\rm r}} ) \ {\rm mod} \ N_1 \right)\cdot p(iT_{\rm s}+\bar{\tau}_{k_r}).
\end{align}
\end{small}
Note that $p(iT_{\rm s})=0$, $i\neq 0$. If $\bar{\tau}_{k_{\rm r}}$ is equal to 0, then $\bar{s}(n_1,\tau_{k_{\rm r}})$ becomes $s((N_1+n_1-l_{k_{\rm r}}) \ {\rm mod} \ N_1)$.
Next, assume that $N_1$ samples are captured at instants $n_2 T_{\rm p}, \dots, n_2 T_{\rm p}+(N_1-1)T_{\rm s}$  and let $\vec{y}_{n_2}=\begin{bmatrix}
y(n_2 T_{\rm p}), \dots, y(n_2 T_{\rm p}+(N_1-1)T_{\rm s})
\end{bmatrix}^T$.  Then, $\vec{y}_{n_2}$ can be written as: 
\begin{equation}
\vec{y}_{n_2}=\vec{S}\vec{\Gamma}\vec{\phi}_{n_2}+ \vec{e}_{n_2}, \label{eq:range}
\end{equation}
where the reflection coefficient matrix $\vec{\Gamma}$ and the steering vector $\vec{\phi}_{n_2}$ are  given by:
\begin{align}
\vec{\Gamma}&=
\left[
 \begin{matrix}
   \gamma_{1,1} & \gamma_{1,2}  & \dots & \gamma_{1,K_{\rm d}}  \\
   \gamma_{2,1} & \gamma_{2,2} & \dots & \gamma_{2,K_{\rm d}}\\
   \vdots        &  \vdots       &  \ddots & \vdots \\   
   \gamma_{K_{\rm r},1}& \gamma_{K_{\rm r},2} & \dots & \gamma_{K_{\rm r},K_{\rm d}} 
  \end{matrix} 
\right]\in \mathbb{C}^{K_{\rm r}\times K_{\rm d}}, \\
\vec{\phi}_{n_2}&=\begin{bmatrix} e^{j\bar{\omega}_1n_2}, & e^{j\bar{\omega}_2n_2},  \dots, & e^{j\bar{\omega}_{K_{\rm d}}n_2}\end{bmatrix}^T,
\end{align}
and $\vec{S}\in \mathbb{R}^{N_1\times K_{\rm r}}$ is defined as: 
\begin{small}
\begin{equation}
\vec{S}=
\left[
 \begin{matrix}
   \bar{s}(0,\tau_{1}) & \bar{s}(0,\tau_{2})  & \dots & \bar{s}(0,\tau_{K_{\rm r}})  \\
   \bar{s}(1,\tau_{1}) & \bar{s}(1,\tau_{2})  & \dots & \bar{s}(1,\tau_{K_{\rm r}}) \\
   \vdots        &  \vdots       &  \ddots & \vdots \\   
   \bar{s}(N_1-1,\tau_{1}) & \bar{s}(N_1-1,\tau_{2})  & \dots & \bar{s}(N_1-1,\tau_{K_{\rm r}}) \\
  \end{matrix}
\right].
\end{equation}
\end{small}
Let \begin{equation}
\vec{Y}=\begin{bmatrix} \vec{y}_0, \vec{y}_1, \dots, \vec{y}_{N_2-1}\end{bmatrix},
\end{equation}
\begin{equation}
\vec{\Phi}=\begin{bmatrix} \vec{\phi}_1,  \vec{\phi}_2,  \dots,  \vec{\phi}_{N_2}\end{bmatrix}^T,
\end{equation}
and 
\begin{equation}
\vec{E}=\begin{bmatrix} \vec{e}_1, \vec{e}_2,  \dots, \vec{e}_{N_2}\end{bmatrix}.
\end{equation}
Then, the observed data matrix can be expressed compactly as follows:
\begin{equation}
\vec{Y}=\vec{S}\vec{\Gamma}\vec{\Phi}^T+\vec{E}. \label{model:LFMCW}
\end{equation}
Let $\vec{\gamma} = {\rm vec}(\vec{\Gamma})$, $\vec{y}={\rm vec}(\vec{Y})$ and $\vec{e}={\rm vec}(\vec{E})$.
Then, the data model in (\ref{model:LFMCW}) can be rewritten as:
\begin{equation}
\vec{y}=\vec{B}\vec{\gamma}+ \vec{e} \in \mathbb{C}^N, \  \vec{B} \in \mathbb{C}^{N\times M},\quad M \gg N, \label{eq:model1}
\end{equation}
where $\vec{B}=\vec{\Phi}\otimes\vec{S}$ is given, $N=N_1N_2$ and $M=K_{\rm r}K_{\rm d}$. Similar to the LFMCW radar case, the range-Doppler imaging problem for PMCW radar can also be treated as a sparse parameter estimation problem with the unknown parameter vector $\vec{\gamma}$ being sparse. 
\subsection{One-Bit Quantization}
\indent Applying one-bit quantization at the receivers of both LFMCW and PMCW radars, we obtain the signed measurement vector $\vec{z}=\vec{z}_{\rm R}+j\vec{z}_{\rm I}$ by comparing the unquantized received signal $\vec{y}$ with a known time-varying threshold $\vec{h}\triangleq \vec{h}_{\rm R}+j\vec{h}_{\rm I}$:
\begin{equation}
    \vec{z}={\rm signc}\left(\vec{B}\vec{\gamma}-\vec{h}+\vec{e}\right), \label{model:one-bit}
\end{equation}
where ${\rm signc}(x)={\rm sign}\left({\rm Re}[x]\right)+j{\rm sign}\left({\rm Im}[x]\right)$  and
\begin{align}
{\rm sign}(x)=\begin{cases} \label{eq:window}
    \ 1 & x \geqslant 0, \\
    -1 & x<0.
    \end{cases}
\end{align}
The problem of interest herein is to find an accurate sparse estimate of $\vec{\gamma}$ given the signed measurement vector $\vec{z}$. In the following sections, we present several approaches to solving  the range-Doppler imaging problem for one-bit automotive radars.  
\section{Overview of the Weighted SPICE framework} \label{section:weighted}
We begin with revisiting  the  SPICE algorithm for the standard high-precision data model in (\ref{eq:linear}) or (\ref{eq:model1}).  To iteratively solve the optimization problem associated with SPICE,  a cyclic approach  was employed in \cite{stoica2010new} and a gradient approach was used  in \cite{stoica2014weighted}. We show that the same weighted  ${\rm SPICE\textsubscript{a}}$ algorithm in \cite{stoica2014weighted} can be derived using an MM approach. The MM derivations of the weighted SPICE algorithms are needed later on when extending them to the one-bit data case. Additionally, we briefly review the basic idea of MM in Appendix \ref{appendix:A}.\\
\indent For notational convenience, we let $\lbrace \gamma_k \rbrace$ denote the elements of the vector $\vec{\gamma}$ in \eqref{eq:linear} or \eqref{eq:model1}.
The SPICE algorithm makes the assumption  that
\begin{equation}
    E\left[ \vec{e}\vec{e}^H\right]=\left[
    \begin{matrix}
       p_{M+1} & 0 & \dots & 0\\
       0 & p_{M+2} & \dots & 0 \\
       \vdots & \vdots & \ddots & \vdots \\
       0 & \dots & \dots &p_{M+N}
    \end{matrix}
    \right],
\end{equation}
and that the phases of $\lbrace\gamma_k\rbrace$ are independently and uniformly
distributed in $[ 0,2\pi]$ (see \cite{stoica2010new} for a discussion showing that these assumptions have only a marginal  effect on the performance of the algorithm). Then,  the covariance matrix of the data vector $\vec{y}$ has the following form:
\begin{align}
    \vec{R}&=E\lbrace \vec{y}\vec{y}^H \rbrace=\vec{B}\vec{P}_1\vec{B}^H+\vec{P}_2 \nonumber \\
    & =\vec{A}\vec{P}\vec{A}^H,
\end{align}
where $\vec{P}_1={\rm diag}(\vec{p}_1)$, with $\vec{p}_1=\begin{bmatrix} p_1,   \dots, p_M \end{bmatrix}^T$ and $p_k=|\gamma_k|^2$; $\vec{P}_2={\rm diag}(\vec{p}_2)$,  with  $\vec{p}_2=\begin{bmatrix} p_{M+1},  \dots,  p_{M+N} \end{bmatrix}^T$; and where
\begin{equation}
    \vec{A}=\begin{bmatrix} \vec{B} & \vec{I}\end{bmatrix}, \quad \vec{P}={\rm diag}(\vec{p}), \quad  \vec{p}=\begin{bmatrix} \vec{p}_1^T &\vec{p}_2^T \end{bmatrix}^T.
\end{equation}
For later use, let $\vec{A}=\begin{bmatrix} \vec{a}_1, \vec{a}_2, \dots, \vec{a}_{M+N}\end{bmatrix}$. 
\subsection{SPICE}
SPICE minimizes the following  covariance fitting criterion \cite{stoica2010new}:
\begin{equation}
    \mathop{\min}_{\vec{p}} \Vert \vec{R}^{-1/2}\left( \vec{y}\vec{y}^H -\vec{R} \right) \Vert_F^2.
\end{equation}
After simple calculations, the SPICE criterion can be reformulated as \cite{stoica2014weighted}:
\begin{align}
    \mathop{\min}_{\Vec{p}} f_1\left(\Vec{p}\right)&= \vec{y}^H\vec{R}^{-1}\vec{y}+ {\rm tr}\lbrace \vec{R} \rbrace \nonumber \\ &=\vec{y}^H\Vec{R}^{-1}\Vec{y}+\sum_{k=1}^{M+N}w_k p_k, \label{criterion:spice}
\end{align}
where $w_k=\Vert\vec{a}_k \Vert^2$. While the optimization problem of minimizing   (\ref{criterion:spice})  is a convex semidefinite program (SDP), the use of a generic SDP solver would be  computationally intensive \cite{stoica2010new}. Using the MM technique, the  optimization problem in (\ref{criterion:spice}) can be efficiently solved iteratively with closed-form solutions  at each iteration. The following   inequality provides an upper bound for  $\vec{R}^{-1}$:
\begin{equation}
    \Vec{R}^{-1}\preceq (\hat{\vec{R}}^t)^{-1}\Vec{A}\hat{\vec{P}}^t\Vec{P}^{-1}\hat{\vec{P}}^t\vec{A}^H(\hat{\vec{R}}^t)^{-1}, \label{ineq:lemma1}
\end{equation}
where $\hat{\vec{P}}^t$ and  $\hat{\vec{R}}^t=\vec{A}\hat{\vec{P}}^t\vec{A}^H$ denote the estimates of $\vec{P}$ and $\vec{R}$ at the $t$th iteration; and the equality is achieved at $\vec{P}=\hat{\vec{P}}^t$. While this property follows from \cite{sun2016robust}, we include a simple proof of it in Appendix \ref{appendix:B} to make the paper self-contained.  Making use of (\ref{ineq:lemma1}), we can obtain a  majorizing  function for  $f_1\left(\Vec{p}\right)$  as follows:
\begin{equation}
    f_1(\vec{p}) \leq g_1(\Vec{p}| \hat{\vec{p}}^t)=(\bar{\vec{\gamma}}^{(t+1)})^H\vec{P}^{-1}\bar{\vec{\gamma}}^{(t+1)}+\sum_{k=1}^{M+N} w_kp_k, 
\end{equation}
where
\begin{align}
    \bar{\vec{\gamma}}^{(t+1)}&=\hat{\vec{P}}^t\vec{A}^H(\hat{\vec{R}}^t)^{-1}\vec{y} \nonumber \\ 
    &\triangleq \begin{bmatrix} ( \hat{\vec{\gamma}}^{(t+1)})^T, \bar{\gamma}^{(t+1)}_{M+1}, \dots, \bar{\gamma}^{(t+1)}_{M+N} \end{bmatrix}^T.
\end{align}
Note that $\hat{\vec{\gamma}}^{(t+1)}=\hat{\vec{P}}_1^t\Vec{B}^H(\hat{\Vec{R}}^t)^{-1}\Vec{y}$ is the  LMMSE estimate of $\vec{\gamma}$  corresponding to the estimate $\hat{\vec{p}}^t_1$ \cite{kay1993,stoica2014weighted}. The majorizing function $g_1(\vec{p}| \hat{\vec{p}}^t)$ is easily minimized  since
\begin{align}
   g_1(\vec{p} | \hat{\vec{p}}^t)&=\sum_{k=1}^{M+N} \frac{\left|\bar{\vec{\gamma}}^{(t+1)}_k\right|^2}{p_k} + w_k p_k  \nonumber \\ 
    &\geqslant \sum_{k=1}^{M+N} 2\sqrt{w_k}\left|\bar{\gamma}^{(t+1)}_k\right|,  \label{eq:1bS}
\end{align}
with the equality achieved for 
\begin{equation}
    \hat{p}^{(t+1)}_k= \left|\bar{\gamma}^{(t+1)}_k\right| / \sqrt{w_k}. \label{eq:power1}
\end{equation}
Inserting $\bar{\gamma}_k^{(t+1)}$ into (\ref{eq:power1})  gives:
\begin{equation}
    \hat{p}_k^{(t+1)}=\hat{p}_k^t\left|\vec{a}_k^H(\hat{\vec{R}}^t)^{-1}\Vec{y}\right|/w_k^{1/2}. \label{eq:SPICEa}
\end{equation}
Given $\hat{\vec{p}}^{(t+1)}$, SPICE updates the convariance matrix via $\hat{\vec{R}}^{(t+1)}=\vec{A}\hat{\vec{P}}^{(t+1)}\vec{A}^H$.  Although derived differently, the updating scheme in (\ref{eq:SPICEa}) coincides  with that of the SPICE\textsubscript{a} algorithm in \cite{stoica2014weighted}. 
The global convergence of the algorithm is guaranteed by  the convexity of the optimization problem and the monotonically decreasing property of the MM technique. 
\par Once $\hat{\vec{p}}$ is obtained, we can estimate $\vec{\gamma}$ via  the well-known LMMSE estimator:  
\begin{equation}
    \hat{\vec{\gamma}}=\hat{\vec{P}}_1\vec{B}^H\hat{\vec{R}}^{-1}\vec{y}.
\end{equation}
\subsection{LIKES} \label{section:likes}
LIKES estimates $\vec{p}$ via minimizing the negative log-likelihood function:
\begin{equation}
    f_2(\Vec{p})=\vec{y}^H\vec{R}^{-1}\Vec{y}+ \ln \left| \Vec{R} \right|. \label{eq:LIKES}
\end{equation}
Note that $\ln \left|\Vec{R} \right|$ is a concave function of $\vec{p}$, and hence the minimization of \eqref{eq:LIKES} is a difficult non-convex optimization problem. We again use the MM technique to  decrease \eqref{eq:LIKES} efficiently at each iteration. A majorizing function for $\ln \left|\Vec{R} \right|$ can be easily constructed by its first-order Taylor expansion:
\begin{equation}
    \ln \left| \vec{R}\right| \leq \sum_{k=1}^{M+N} \left(\vec{a}_k^H(\hat{\Vec{R}}^t)^{-1}\Vec{a}_k\right)p_k + {\rm const}. \label{ineq:LIKES}
\end{equation}
Using   (\ref{ineq:LIKES}) yields the following majorizing function for $f_2(\vec{p})$:
\begin{equation}
    f_2(\Vec{p}) \leq g_2(\Vec{p} | \hat{\vec{p}}^t)=\vec{y}^H\vec{R}^{-1}\Vec{y} + \sum_{k=1}^{M+N}w_kp_k, 
\end{equation}
with 
\begin{equation}
    w_k=\vec{a}_k^H(\hat{\vec{R}}^t)^{-1}\vec{a}_k. \label{weights:LIKES}
\end{equation}
Different from the constant weights of the SPICE algorithm,  the weights of LIKES are updated during the MM  iterations. For fixed weights at the $(t+1)$th iteration, the majorizing function $g_2(\vec{p}| \hat{\vec{p}}^t)$ has the same form as the  SPICE criterion (see (\ref{criterion:spice})). Hence the decrease or minimization of $g_2(\vec{p}| \hat{\vec{p}}^t)$ can be achieved via the SPICE solver presented in the previous subsection. LIKES consists of the inner and outer iterations. At the $(t+1)$th outer iteration,
\begin{align}
    \hat{p}_{l+1,k}^{(t+1)}&=\hat{p}_{l,k}^{(t+1)} \left|\vec{a}_k^H\left(\hat{\vec{R}}_{l}^{(t+1)}\right)^{-1}\vec{y}\right|/ \left(\Vec{a}_k^H(\hat{\Vec{R}}^t)^{-1}\Vec{a}_k\right)^{1/2}, \nonumber \\
    \quad  l&=0,\dots, L-1, \label{power:likes}
\end{align}
where $\hat{p}_{l,k}^{(t+1)}$ and $\hat{\vec{R}}_l^{(t+1)}$ denote the estimates of $p_k$ and $\vec{R}$, respectively, at the $l$th inner iteration and the $(t+1)$th outer iteration, with $\hat{p}_{0,k}^{(t+1)}=\hat{p}^{t}_k$ and $\hat{\vec{R}}^{(t+1)}_0=\hat{\vec{R}}^t$; and $L$ denotes the number of  inner iterations.  In practice, $L$ can be a small number or even just $1$.  After $L$ inner iterations, we let $\hat{p}_k^{(t+1)}=\hat{p}_{L,k}^{(t+1)}$ and $\hat{\vec{R}}^{(t+1)}=\hat{\vec{R}}_L^{(t+1)}$, and update the weights $\lbrace w_k \rbrace$ according to (\ref{weights:LIKES}) and  continue with the next outer iteration. LIKES  converges locally,  but the global convergence of the algorithm is  difficult to analyze due to the fact that \eqref{eq:LIKES} is a non-convex function of $\vec{p}$.  
\subsection{SLIM} \label{section:SLIM}
By making use of a slightly different form of the  sparsity enforcing term of LIKES, the criterion optimized by SLIM has the following form:
\begin{equation}
    f_3(\Vec{\Vec{p}})=\Vec{y}^H\Vec{R}^{-1}\Vec{y}+\ln \left| \Vec{P}\right|. \label{eq:SLIM}
\end{equation}
Again, $\ln \left|\Vec{P} \right|$ is a concave function of $\vec{p}$ and a majorizing function can be constructed by a first-order Taylor expansion: 
\begin{equation}
    \ln \left| \Vec{P} \right|= \sum_{k=1}^{M+N} \ln \left| p_k \right| \leq \sum_{k=1}^{M+N}\frac{p_k}{\hat{p}^t_k} +{\rm const}.  \label{ineq:SLIM}
\end{equation}
Using  (\ref{ineq:lemma1}) and (\ref{ineq:SLIM}) yields the following majorizing function for $f_3(\Vec{p})$:
\begin{equation}
        f_3(\Vec{p}) \leq g_3(\Vec{p}| \hat{\vec{p}}^t)=(\bar{\Vec{\gamma}}^{(t+1)})^H\Vec{P}^{-1}\bar{\Vec{\gamma}}^{(t+1)} + \sum_{k=1}^{M+N}w_kp_k,  \label{weights:slim}
\end{equation}
where $w_k=1/{\hat{p}_k^t}$. Minimizing $g_3(\vec{p}| \hat{\vec{p}}^t)$ yields the $(t+1)$th updating formula of SLIM:
\begin{equation}
    \hat{p}^{(t+1)}_k=(\hat{p}_k^t)^{\frac{3}{2}}\left|\vec{a}^H(\hat{\vec{R}}^t)^{-1}\Vec{y}\right|. \label{power:slim}
\end{equation}
Similar  to LIKES, SLIM monotonically decreases the objective function in (\ref{eq:SLIM}) and  converges locally.
\subsection{IAA}
There is no known objective function-based derivation of IAA that is similar to (\ref{criterion:spice}), (\ref{eq:LIKES}) or (\ref{eq:SLIM})  above. According to the approximate interpretation of this algorithm in \cite{yardibi2010source,roberts2010iterative}, the weight of IAA is given by:
\begin{equation}
    w_k=\hat{p}_k^t\left(\vec{a}^H_k(\hat{\Vec{R}}^t)^{-1}\vec{a}_k\right)^2. \label{weight:IAA}
\end{equation}
Inserting (\ref{weight:IAA}) in (\ref{eq:SPICEa}) yields the IAA updating formula:
\begin{equation}
    \hat{p}_k^{(t+1)}=(\hat{p}_k^t)^{1/2}\left|\vec{a}_k^H(\hat{\vec{R}}^t)^{-1}\Vec{y}\right|/ \left(\Vec{a}_k^H(\hat{\Vec{R}}^t)^{-1}\Vec{a}_k\right).
\end{equation}
All  four  algorithms above can be unified under a  weighted SPICE framework (as mentioned above,  IAA can be cast in this framework only approximately):
\begin{equation}
    \mathop{\min}_{\vec{p}}\vec{y}^H\vec{R}^{-1}\vec{y} + \sum_{k=1}^{M+N} w_kp_k.
\end{equation}
The different choices of the weights $\lbrace w_k \rbrace$ in  (\ref{criterion:spice}), (\ref{weights:LIKES}), (\ref{weights:slim}) and (\ref{weight:IAA}) lead to  four different hyperparameter-free algorithms. SPICE has constant weights, while LIKES, SLIM and IAA use adaptive weights and can be interpreted as adaptively reweighted SPICE methods.
\section{One-bit SLIM} \label{sec:one-bit weighted spice}
In this section,  we present a regularized minimization approach, referred to as 1bSLIM, for the one-bit case.  After one-bit quantization, the linear model (\ref{eq:linear}) or (\ref{eq:model1}) becomes the nonlinear model (\ref{model:one-bit}). The problem is to estimate $\vec{\gamma}$ from the one-bit quantized signal $\vec{z}$, making use of the fact that the time-varying threshold $\vec{h}$ is known. \\
\indent Assume that $\vec{e}$ is i.i.d.  circularly symmetric complex-valued white Gaussian noise with zero-mean and unknown variance $\sigma^2$, i.e., ${\rm Re}\left[ e(n)\right]\sim \mathcal{N}(0,\frac{\sigma^2}{2 })$ and ${\rm Im}\left[ e(n)\right]\sim \mathcal{N}(0,\frac{\sigma^2}{2 })$. Then, the  negative log-likelihood function of the signed measurement vector is given by \cite{gianelli2016one,ren2019sinusoidal}:
\begin{align}
L(\vec{\gamma},\sigma)=-\sum_{n=1}^N\ln \left(\Phi\left(z_{\rm R}(n)\frac{{\rm Re}\left[\vec{b}_n^T\vec{\gamma}\right]  -h_{\rm R}(n)}{\sigma / \sqrt{2}} \right) \right) \nonumber \\
  -\sum_{n=1}^N\ln\left(\Phi\left(z_{\rm I}(n)\frac{{\rm Im}\left[\vec{b}_n^T\vec{\gamma}\right] -{h}_{\rm I}(n) }{\sigma / \sqrt{2}} \right) \right), \label{eq:ml}
\end{align}
where $\Phi(x)=\frac{1}{\sqrt{2\pi}}\int_{-\infty}^{x}e^{-\frac{t^2}{2}}dt$ is the cumulative density function of the standard Gaussian distribution, and  $\vec{b}_n$ is the $n$th column of $\vec{B}^T$. For convenience, we reparameterize the negative log-likelihood function by defining $\eta=\frac{\sqrt{2}}{\sigma}$ and  $\vec{\beta}=\eta\vec{\gamma}$. Then, (\ref{eq:ml}) can be reformulated as: 
\begin{align}
L(\vec{\beta},\eta)=&-\sum_{n=1}^N\ln\left(\Phi\left(z_{\rm R}(n)\left({\rm Re}\left[\vec{b}_n^T\vec{\beta}\right]  -\eta h_{\rm R}(n) \right) \right) \right)  \nonumber \\
  &-\sum_{n=1}^N\ln\left(\Phi\left(z_{\rm I}(n)\left({\rm Im}\left[\vec{b}_n^T\vec{\beta}\right] -\eta{h}_{\rm I}(n) \right) \right) \right). \label{eq:surrogate}
\end{align}
Consider minimizing  the following  regularized negative log-likelihood function for sparse parameter estimation:
\begin{equation}
     G(\vec{\beta},\eta)=L(\vec{\beta},\eta) + \sum_{k=1}^M{\ln{ (|\beta_k|^2+\varepsilon) } }, \label{criterion:1bLSIMa}
\end{equation}
where $\varepsilon >0$ is a small positive number  that makes sure that the function has a finite lower bound. The first term of $G(\vec{\beta},\eta)$ (i.e., $L(\vec{\beta},\eta)$) is the fitting term and the second term (i.e., $\sum_{k=1}^M{\ln{ (|\beta_k|^2+\varepsilon)} }$) is a sparsity enforcing  term. Similar to the original SLIM algorithm,  1bSLIM can be interpreted as a MAP approach.  Consider the following Bayesian model:
\begin{align}
    p(\vec{z}| \vec{\beta},\eta) &=\prod_{n=1}^N \Phi\left( z_{\rm R}(n)\left({\rm Re}\left[\vec{b}_n^T\vec{\beta}\right]  -\eta h_{\rm R}(n) \right) \right) \nonumber \\ &\qquad\quad\Phi\left(z_{\rm I}(n)\left({\rm Im}\left[\vec{b}_n^T\vec{\beta}\right] -\eta{h}_{\rm I}(n) \right) \right),  \nonumber \\
    p(\vec{\beta}) &\propto \prod_{k=1}^M \frac{1}{|\beta_k|^2+\varepsilon}, \  p(\eta) \propto 1,
\end{align}
where $p(\vec{\beta}) \propto \prod_{k=1}^M \frac{1}{|\beta_k|^2+\varepsilon}$ is a  sparsity promoting prior,  and $p(\eta) \propto 1$ is a  noninformative  prior \cite{gelman2013bayesian}. We can estimate $\vec{\beta}$ and $\eta$ via the MAP approach:
\begin{equation}
    \left(\vec{\beta},\eta \right)=\arg \mathop{\max}_{\vec{\beta},\eta} p(\vec{z}| \vec{\beta},\eta)p(\vec{\beta})p(\eta). \label{eq:MAP}
\end{equation}
Note that after taking the negative logarithm of the above expression,  (\ref{eq:MAP}) becomes equivalent to (\ref{criterion:1bLSIMa}).  
\par To efficiently minimize the complicated and non-convex objective function in (\ref{criterion:1bLSIMa}), we utilize the MM approach once again. Specifically, we first derive  a majorizing function for $L(\vec{\beta},\eta)$. After that,   a closed-form updating  formula for each MM iteration is obtained. 
\subsection{Majorizing Function for $L(\vec{\beta},\eta)$}
Let $f(x)=-\ln \left( \Phi(x)\right)$. The following inequality holds  for any $u,x\in \mathbb{R}$ \cite{ren2019sinusoidal}:
\begin{equation}
f(x)\leq f(u)+f'(u)(x-u)+ \frac{1}{2}(x-u)^2. \label{eq:MM}
\end{equation}  
Let 
\begin{align}
x_{\rm R}(n)&=z_{\rm R}(n)\left({\rm Re}\left[\vec{b}_n^T\vec{\beta}\right]-\eta h_{\rm R}(n)\right), \nonumber \\
x_{\rm I}(n)&=z_{\rm I}(n)\left({\rm Im}\left[\vec{b}_n^T\vec{\beta}\right]-\eta h_{\rm I}(n)\right), \nonumber \\
f'(x)&=-\frac{\phi(x)}{\Phi(x)}, \nonumber
\end{align}
where $\phi(x)=\frac{1}{\sqrt{2\pi}}e^{-\frac{x^2}{2}}$ is  the standard normal probability density function. Then, the negative log-likelihood function $L(\vec{\beta},\eta)$ in (\ref{eq:surrogate}) can be written as:
\begin{equation}
L(\vec{\beta},\eta)=\sum_{n=1}^N f\left(x_{\rm R}(n)\right)+\sum_{n=1}^N f\left(x_{\rm I}(n)\right).
\end{equation}
Assuming  that  $\hat{\vec{\beta}}^t$ and $\hat{\eta}^t$ are available from the $t$th iteration, we can obtain a majorizing  function for $L(\vec{\beta},\eta)$ by making use of  (\ref{eq:MM}):
\begin{align}
L(\vec{\beta},\eta)&\leq  \sum_{n=1}^N\frac{1}{2} x_{\rm R}^2(n)+\frac{1}{2}x_{\rm I}^2(n) \nonumber\\
&- u^t_{\rm R}(n)x_{\rm R}(n)- u^t_{\rm I}(n)x_{\rm I}(n) + {\rm const}, \label{eq:mm}
\end{align}
where $u^t_{\rm R/I}(n)$ is given by: 
\begin{equation}
u_{\rm R/I}^t(n)=x_{\rm R/I}^t(n)-f'\left(x_{\rm R/I}^t(n)\right).
\end{equation}
A simple calculation shows that (\ref{eq:mm}) can be re-written as follows: 
\begin{equation}
L(\vec{\beta},\eta) \leq   \frac{1}{2}\Vert \vec{B}\vec{\beta}-\left(\eta\vec{h}+\vec{g}^t\right)\Vert^2+{\rm const}, \label{eq:L}
\end{equation}
where $\vec{g}^t\!=\!\begin{bmatrix}  g^t(1), \dots, g^t(N) \end{bmatrix}^T$, with
\begin{equation}
    g^t(n)= z_{\rm R}(n)u_{\rm R}^t(n)+jz_{\rm I}(n)u_{\rm I}^t(n).   \label{eq:tmp}
\end{equation}
\subsection{Updating Formula}
Making use of  (\ref{eq:L}) and (\ref{ineq:SLIM})  and ignoring the terms independent of $\vec{\beta}$ and $\eta$ yield the following majorizing function for the objective function $G(\vec{\beta},\eta)$ at the $(t+1)$th iteration:
\begin{equation}
Q(\vec{\beta},\eta | \hat{\vec{\beta}}^t,\hat{\eta}^t)=
\frac{1}{2}\Vert  \vec{B}\vec{\beta}-\left(\eta\vec{h}+\vec{g}^t\right)\Vert^2 
+ \vec{\beta}^H\left(\hat{\vec{P}}^t\right)^{-1}\vec{\beta} ,   \label{eq:quadratic}
\end{equation}
where $\hat{\vec{P}}^t={\rm diag} \lbrace \hat{\vec{p}}^t \rbrace$ with $\hat{\vec{p}}^t=\begin{bmatrix} \hat{p}^t_1,\hat{p}^t_2,\dots, \hat{p}_M^t \end{bmatrix}$ and
\begin{equation}
    \hat{p}^{t}_k=|\hat{\beta}_k^t|^2 +\varepsilon. \label{power:1bslim}
\end{equation}
Note that minimizing $Q\left(\vec{\beta},\eta| \hat{\vec{\beta}}^t,\hat{\eta}^t \right)$ with respect to $\vec{\beta}$ and $\eta$ is an unconstrained quadratic optimization problem, and we can set the complex derivatives $\left(d/d\vec{\beta}^H\right)Q(\vec{\beta},\eta | \hat{\vec{\beta}}^t,\hat{\eta}^t)$ and $\left(d/d\eta\right)Q(\vec{\beta},\eta| \hat{\vec{\beta}}^t, \hat{\eta}^t)$ to zeros to obtain $\hat{\vec{\beta}}^{(t+1)}$ and $\hat{\eta}^{(t+1)}$ (ignoring, for a moment, the fact that $\eta \geqslant 0$). This approach leads to the following linear equations:
\begin{align}
\left(\vec{B}^H\vec{B}+2\left(\hat{\vec{P}}^t\right)^{-1}\right)\vec{\beta}-\eta\vec{B}^H \vec{h}&=\vec{B}^H\vec{g}^t,  \label{eq:beta2} \\
-{\rm Re}\left[\vec{h}^H\vec{B}\vec{\beta}\right]+\eta\vec{h}^H\vec{h}&=-{\rm Re}\left[\vec{h}^H\vec{g}^t\right]. \label{eq:solution}
\end{align}
Since the matrix  $\vec{B}^H\vec{B}+2\left(\hat{\vec{P}}^t\right)^{-1}$ is positive definite, we can obtain $\vec{\beta}$ from Equation (\ref{eq:beta2}) as follows:
\begin{align}
\vec{\beta}&=\left[\vec{B}^H\vec{B}+2\left(\hat{\vec{P}}^t\right)^{-1}\right]^{-1}\vec{B}^H(\eta\vec{h}+\vec{g}^t), \nonumber \\
&=\hat{\vec{P}}^t\vec{B}^H(\hat{\vec{R}}^t)^{-1}\left(\eta\vec{h}+\vec{g}^t\right), \label{eq:eliminate}
\end{align}
where
\begin{equation}
    \hat{\vec{R}}^t=\vec{B}\hat{\vec{P}}^t\vec{B}^H+2\vec{I}. \label{eq:R}
\end{equation}
 Substituting  $\vec{\beta}$ from (\ref{eq:eliminate}) into  (\ref{eq:solution}), we  obtain the solution for $\hat{\eta}^{\left(t+1\right)}$:
\begin{equation}
\hat{\eta}^{(t+1)}=\max \left(0, -\frac{{\rm Re}\left[\vec{h}^H(\hat{\vec{R}}^t)^{-1}\vec{g}^t\right]}{\vec{h}^H(\hat{\vec{R}}^t)^{-1}\vec{h}}\right). \label{eq:eta}
\end{equation}
Finally, substituting  $\hat{\eta}^{(t+1)}$ into (\ref{eq:eliminate}) we get: 
\begin{equation}
\hat{\vec{\beta}}^{(t+1)}=\hat{\vec{P}}^t\vec{B}^H(\hat{\vec{R}}^t)^{-1}\left(\hat{\eta}^{(t+1)}\vec{h}+\vec{g}^t\right). \label{eq:beta}
\end{equation}
By making use of the MM technique, the majorizing function \eqref{eq:quadratic}, which  we derived above, is much easier to minimize than the original objective function \eqref{criterion:1bLSIMa}. The objective function $G(\vec{\beta},\eta)$ is guaranteed to decrease monotonically due to the MM methodology used to derive the algorithm, and hence 1bSLIM converges to a local minimum of the cost function in (\ref{criterion:1bLSIMa}). 
\section{one-bit weighted SPICE framework} \label{Section:one-bit weighted}
In this section, we first discuss the relationship between the proposed 1bSLIM for one-bit data and the original SLIM algorithm for high-precision data. Using the MM approach, each iteration of the 1bSLIM can be interpreted as a step of the original SLIM applied to a modified high-precision data, and has an updating formula similar to that of the original SLIM. Inspired by this connection and the MM derivations of the weighted SPICE algorithms, we also extend SPICE, LIKES and IAA to the one-bit data case.
\subsection{The Connection Between 1bSLIM and SLIM} \label{section:connection}
Consider the following cost function:
\begin{equation}
\bar{G}(\vec{\beta},\vec{p},\eta)=L(\vec{\beta},\eta)+ \sum_{k=1}^{M} \left|\beta_k \right|^2/ p_k + \sum_{k=1}^{M}\ln\left( p_k+\varepsilon\right). \label{eq:Ga}
\end{equation}
Note that minimizing $\bar{G}(\vec{\beta},\vec{p},\eta)$ with respect to $\vec{p}$ (for $\varepsilon=0$) gives $p_k=|\beta_k|^2$ and hence \eqref{eq:Ga} reduces to \eqref{criterion:1bLSIMa}. This fact shows that (\ref{eq:Ga}) is an augmented form of (\ref{criterion:1bLSIMa}) and  hence the minimization of \eqref{eq:Ga} can be conveniently done by means of minimizing \eqref{criterion:1bLSIMa} with respect to $\vec{\beta}$ and $\eta$. 
\par In this section, we use  \eqref{eq:Ga} to discuss the connection between 1bSLIM and SLIM. Making use of the inequality (\ref{ineq:SLIM}) and (\ref{eq:L})  and  ignoring the terms independent of $\vec{\beta}$, $\vec{p}$ and $\eta$ yield the following majorizing function for the objective function $\bar{G}(\vec{\beta},\vec{p},\eta)$:
\begin{equation}
\bar{Q}(\vec{\beta},\vec{p},\eta)=
\frac{1}{2}\Vert  \vec{B}\vec{\beta}-\left(\eta\vec{h}+\vec{g}^t\right)\Vert^2 
+ \vec{\beta}^H\vec{P}^{-1}\vec{\beta} +\sum_{k=1}^{M}w_k p_k,  \label{eq:quadratic2}
\end{equation}
where
\begin{equation}
   \vec{P}={\rm diag}\lbrace \vec{p} \rbrace \ {\rm and} \  w_k=1 / \left( \hat{p}_k^t +\varepsilon\right). \label{w:slim}
\end{equation}
Let  
\begin{equation}
    \bar{\vec{y}}^{(t+1)}=\hat{\eta}^{(t+1)}\vec{h}+\vec{g}^t. \label{eq:infi_data}
\end{equation}
Then, $\bar{Q}\left(\vec{\beta},\vec{p},\hat{\eta}^{(t+1)}\right)$ becomes
\begin{equation}
    \bar{Q}\left(\vec{\beta},\vec{p},\hat{\eta}^{(t+1)}\right)=\frac{1}{2}\Vert \bar{\vec{y}}^{(t+1)}-\vec{B}\vec{\beta} \Vert^2 +\sum_{k=1}^{M}\frac{\left|\beta_k\right|^2}{p_k}+ \sum_{k=1}^M w_k p_k. \label{eq:connection}
\end{equation}
The minimization of $\bar{Q}\left(\vec{\beta},\vec{p},\hat{\eta}^{(t+1)}\right)$ with respect to $\vec{\beta}$ yields:
\begin{equation}
    \min_{\vec{\beta}}\bar{Q}\left(\vec{\beta},\vec{p},\hat{\eta}^{(t+1)}\right)= \left(\bar{\vec{y}}^{(t+1)}\right)^H\vec{R}^{-1}\bar{\vec{y}}^{(t+1)}+\sum_{k=1}^M w_k p_k, \label{eq:con1}
\end{equation}
where $\vec{R}=\vec{B}\vec{P}\vec{B}^H+2\vec{I}$, and  the minimum is achieved at:
\begin{equation}
    \vec{\beta}=\vec{P}\vec{B}^H\vec{R}^{-1}\bar{\vec{y}}^{(t+1)}.
\end{equation}
Hence, $\bar{Q}(\vec{\beta},\vec{p},\hat{\eta}^{(t+1)})$ is an augmented function for the weighted SPICE criterion in (\ref{eq:con1}). By interpreting  $\bar{\vec{y}}^{(t+1)}$  as a modified high-precision data, the weighted SPICE criterion in (\ref{eq:con1}) coincides with that of the  original SLIM. 
\par In sum, by using the MM technique the sparse parameter estimation problem for the nonlinear model in (\ref{model:one-bit}) can be approximately solved by making use of a linear model with a  high-precision data $\bar{\vec{y}}^{(t+1)}$ given by:
\begin{equation}
    \bar{\vec{y}}^{(t+1)}=\vec{B}\vec{\beta} + \bar{\vec{e}}, \label{eq:modified}
\end{equation}
where the mean and covariance matrix of the noise vector  $\bar{\vec{e}}$ are $0$ and $2\vec{I}$, respectively. Then, the amplitude $\vec{\beta}$ and power $\vec{p}$ of 1bSLIM can be equivalently obtained by applying SLIM to $\bar{\vec{y}}^{(t+1)}$ in (\ref{eq:modified}).
\subsection{Extending  SPICE, LIKES and IAA to the One-Bit Case}
Section \ref{section:weighted} showed  that  four different choices of the weight of the weighted SPICE algorithm correspond to four different sparsity enforcing terms and they lead to  four different hyperparameter-free sparse parameter estimation algorithms.  
We now use the sparsity enforcing terms of SPICE, LIKES and IAA to obtain extensions of these algorithms to the one-bit data case. 
\subsubsection{1bSPICE}
Replacing the sparsity enforcing term in \eqref{eq:Ga}  by ${\rm tr}\lbrace\vec{R} \rbrace$, we get the following optimization criterion:
\begin{equation}
    G_1(\vec{\beta},\vec{p},\eta)=L(\vec{\beta},\eta)+\sum_{k=1}^M \left| \beta_k \right| ^2 / p_k + {\rm tr} \lbrace \vec{R} \rbrace.  \label{criterion:1bSPICE}
\end{equation}
Making use of \eqref{eq:L} yields the following majorizing function for \eqref{criterion:1bSPICE}:
\begin{equation}
\bar{Q}_1(\vec{\beta},\vec{p},\eta)=
\frac{1}{2}\Vert  \vec{B}\vec{\beta}-\left(\eta\vec{h}+\vec{g}^t\right)\Vert^2 
+ \vec{\beta}^H\vec{P}^{-1}\vec{\beta} +\sum_{k=1}^{M}w_k p_k,\label{eq:quadspice}
\end{equation}
where $ w_k=\Vert \vec{b}_k \Vert^2$. The minimization of  (\ref{eq:quadspice}) can be  achieved by a blockwise  cyclic  algorithm, which alternatingly minimizes (\ref{eq:quadspice}) with respect to $\lbrace\Vec{\beta},\eta\rbrace$ for fixed $\vec{p}$ and with respect to $\vec{p}$ for fixed $\lbrace \vec{\beta},\eta \rbrace$. That is, $\lbrace\Vec{\beta},\eta\rbrace$ and $\vec{p}$ are updated iteratively by solving the following subproblems:
\begin{align}
    &\lbrace\hat{\vec{\beta}}^{(t+1)},\hat{\eta}^{(t+1)}\rbrace = \arg \mathop{\min}_{\Vec{\beta},\eta} \bar{Q}_1\left(\Vec{\beta},\hat{\vec{p}}^t,\eta \right), \label{eq:cyclic1} \\
    &\hat{\vec{p}}^{(t+1)}= \arg \mathop{\min}_{\vec{p}} \bar{Q}_1\left(\hat{\Vec{\beta}}^{(t+1)}, \vec{\vec{p}},\hat{\eta}^{(t+1)}\right).  \label{eq:cyclic}
\end{align}
Note that the optimization problem in \eqref{eq:cyclic1} is similar to the one in   (\ref{eq:quadratic}).  Hence the updating formulas for $\hat{\eta}^{(t+1)}$ and $\hat{\vec{\beta}}^{(t+1)}$ are similar to those in (\ref{eq:eta}) and (\ref{eq:beta}), respectively. The solution to the optimization problem in  \eqref{eq:cyclic} is:
\begin{equation}
    \hat{p}^{(t+1)}_k= \left|\hat{\beta}^{(t+1)}_k\right| / \sqrt{w_k}. \label{eq:power}
\end{equation}
Extensive numerical examples showed that decreasing \eqref{eq:quadspice} by iterating \eqref{eq:cyclic1} and \eqref{eq:cyclic} only once can typically provide a similar performance to that corresponding to multiple iterations of these two equations, but at a reduced overall computational cost.  The monotonicity  property of  the blockwise   cyclic  algorithm  is guaranteed  since
\begin{align}
    \bar{Q}_1\left(\hat{\vec{\beta}}^t, \hat{\vec{p}}^t,\hat{\eta}^t\right) &\geqslant   \bar{Q}_1\left(\hat{\vec{\beta}}^{(t+1)},\hat{\vec{p}}^t, \hat{\eta}^{(t+1)}\right)  \nonumber  \\ &\geqslant  \bar{Q}_1\left(\hat{\vec{\beta}}^{(t+1)},\hat{\vec{p}}^{(t+1)}, \hat{\eta}^{(t+1)}\right).  \nonumber \label{ineq:converge}
\end{align}
The first inequality comes from the minimization of   $\bar{Q}_1(\vec{\beta},\hat{\vec{p}}^t,\eta)$ with respect to $\lbrace \vec{\beta},\eta \rbrace$, and the second inequality can be guaranteed via 
decreasing $\bar{Q}_1(\hat{\vec{\beta}}^{(t+1)}, \vec{p}, \hat{\eta}^{(t+1)})$ with respect to $\vec{p}$. 
\par We refer to the above algorithm as 1bSPICE due to its relationship to the SPICE algorithm. 1bSPICE decreases (\ref{criterion:1bSPICE}) at each iteration and converges to a local minimum. \\

\subsubsection{1bLIKES}
Replacing the sparsity enforcing term in \eqref{eq:Ga} by $\ln{|\vec{R}|}$, we obtain the following criterion:
\begin{equation}
        G_2(\vec{\beta},\vec{p},\eta)=L(\vec{\beta},\eta)+\sum_{k=1}^M \left| \beta_k \right| ^2 / p_k + \ln{\left|\vec{R}\right|}.  \label{criterion:1blikes}
\end{equation} 
By making use of \eqref{ineq:LIKES} and \eqref{eq:L}, we can obtain  a majorizing function for $G_2(\vec{\beta},\vec{p},\eta)$ having  the same form as the one  in \eqref{eq:quadspice} (except that the weight $w_k$ in \eqref{eq:quadspice} takes the form in \eqref{weights:LIKES}). The updating formulas for $\eta, \vec{\beta}$ and $\vec{p}$ have similar  forms as those  in \eqref{eq:eta}, \eqref{eq:beta} and \eqref{eq:power}. 
We refer to the resulting algorithm as 1bLIKES, because 1bLIKES is related to LIKES algorithm through \eqref{eq:connection} and \eqref{eq:con1}. 1bLIKES  generates a sequence of estimates that monotonically decreases \eqref{criterion:1blikes} and converges to a local minimum of this function. By Lemma 1 in \cite{stoica2014weighted}:
 \begin{equation}
    \vec{b}_k(\hat{\vec{R}}^t)^{-1}\vec{b}_k \leq \frac{1}{\hat{p}_k^t}, \label{ineq:lemma2}
 \end{equation}
which implies that the weight of 1bLIKES is smaller than that of 1bSLIM (for $\varepsilon=0$). Consequently 1bLIKES will give less sparse results than 1bSLIM. 
\begin{table*}[htb]
    \renewcommand\arraystretch{1.7}
    \centering
    \caption{ The one-bit weighted SPICE algorithms}
    \label{Table:table1}
\begin{tabular}{|l|c|p{8cm}|c|}
\hline 
\qquad \qquad Step & 
\multicolumn{2}{c|}{Computation/operation} & Eq. no\\
\hline
1. Initialization & \multicolumn{2}{c|}{$\hat{\beta}_k^0=(1+j)\times 10^{-3}$, \ $\hat{p}_k^0=|\hat{\beta}_k^0|^2$} & \\
\hline 
2. Computation of $\vec{g}^t$ & \multicolumn{2}{c|}{$g^t(n)= z_R(n)u_R^t(n)+jz_I(n)u_I^t(n)$} &  (\ref{eq:tmp}) \\
\hline
3. Computation of $\hat{\vec{R}}^t$ & \multicolumn{2}{c|}{$\hat{\vec{R}}^t=\vec{B}\hat{\vec{P}}^t\vec{B}^H + 2\vec{I}$} &  (\ref{eq:R}) \\
\hline
\multirow{4}{*}{4. Update of weights} &
  1bSPICE & \qquad \quad $w_k=\Vert \vec{b}_k \Vert^2$  & (\ref{criterion:spice}) \\ 
\cline{2-4}
& 1bLIKES & \qquad \quad $w_k=\vec{b}_k^H(\hat{\vec{R}}^t)^{-1}\vec{b}_k$  &  (\ref{weights:LIKES})\\
\cline{2-4}
& \hspace{-12pt}1bIAA & \qquad \quad $w_k=\hat{p}_k^t\left(\vec{b}^H_k(\hat{\Vec{R}}^t)^{-1}\vec{b}_k\right)^2$ & (\ref{weight:IAA})\\
\hline 
5. Computation of $\hat{\eta}^{(t+1)}$  &  
\multicolumn{2}{l|}{\qquad $\hat{\eta}^{(t+1)}=\max \left(0, -{\rm Re}\left[\vec{h}^H(\hat{\vec{R}}^t)^{-1}\vec{g}^t\right] /    {\vec{h}^H(\hat{\vec{R}}^t)^{-1}\vec{h}} \right) $}  & (\ref{eq:eta}) \\
\hline
6. Computation of $\hat{\vec{\beta}}^{(t+1)}$  & 
\multicolumn{2}{l|}{\qquad $\hat{\vec{\beta}}^{(t+1)}=\hat{\vec{P}}^t\vec{B}^H(\hat{\vec{R}}^t)^{-1}\left(\hat{\eta}^{(t+1)}\vec{h}+ \vec{g}^t\right)$}   & (\ref{eq:beta})\\
\hline
\multirow{2}{*}{7. Update of power}&
1bSLIM & \qquad $\hat{p}_k^{(t+1)}=|\hat{\beta}^{(t+1)}_k|^2 +\varepsilon, \quad k=1,\dots,M$  & (\ref{power:1bslim})\\
\cline{2-4} 
& Others & \qquad $\hat{p}_k^{(t+1)}=\left|\hat{\beta}^{(t+1)}_k\right| / \sqrt{w_k}, \quad k=1,\dots,M$    & (\ref{eq:power})\\
\hline
\multicolumn{4}{|c|}{ Iterate Steps 2--7 until practical  convergence or $t$ reaches the maximum number.}  \\
\hline
\end{tabular}
\end{table*} \\
\subsubsection{1bIAA}
Finally, replacing the weight in  (\ref{eq:quadspice}) by the IAA's weight in (\ref{weight:IAA}) yields an augmented function for the IAA criterion, and we refer to the resulting  algorithm as 1bIAA. From (\ref{ineq:lemma2}), we note that the IAA's weights are smaller than those of LIKES and SLIM. Hence it is conceivable that the results obtained by 1bIAA are less sparse than those obtained with both 1bSLIM and 1bLIKES. The updating formulas of 1bIAA are the same as those of 1bSPICE with the exception of replacing $w_k$ with the IAA's weights in (\ref{weight:IAA}).  Although we were unable to find an optimization metric for 1bIAA,  we have never encountered an example where 1bIAA did  not converge (the study of the convergence properties of both IAA and  1bIAA is an interesting topic for future research).
\par \textit{Remark 1:} The majorizing functions of 1bSPICE, 1bLIKES, 1bSLIM and 1bIAA at the $(t+1)$th iteration  have  the same form as in (\ref{eq:con1})  but with different weights. Hence these four algorithms can be unified under the one-bit weighted SPICE umbrella.  The one-bit weighted SPICE is related to the conventional weighted SPICE in that  the former  is equivalent to applying the latter at the $(t+1)$th iteration to the modified high-precision data  in (\ref{eq:modified}). The  proposed  one-bit  weighted  SPICE  framework  is  summarized  in  Table \ref{Table:table1}. 

\subsection{Implementations} \label{section: implementations}
\subsubsection{Initialization and termination}
\par  The proposed four algorithms are initialized  with  some small complex-valued amplitude estimates, e.g., equal to $(1+j)\times10^{-3}$.  ``Practical convergence'' is achieved when the relative change of $\vec{p}$ between two consecutive iterations is less than $10^{-3}$ or a maximum iteration number equal to 150 is reached. $\varepsilon$ in (\ref{criterion:1bLSIMa}) is typically chosen as a small positive number (e.g., $\varepsilon=10^{-4}$).   
\subsubsection{Fast implementations of 1bSLIM and 1bSPICE}
In the  iterations of 1bSLIM and 1bSPICE,  we need to compute the covariance matrix $\hat{\vec{R}}$ and its inverse from the power estimate $\hat{\vec{P}}$ obtained from the previous iteration,  an operation that  has a high computational burden.  We can reduce the computational burden via first calculating $\vec{v}_1^t=(\hat{\vec{R}}^t)^{-1}\vec{h}$ and $\vec{v}_2^t=(\hat{\vec{R}}^t)^{-1}\vec{g}^t$ using the CGLS approach \cite{daniel,tan2010sparse}. Then, $\hat{\eta}^{(t+1)}=-\max \lbrace 0,{\rm Re}\left[\vec{h}^H\vec{v}_2^t\right] / \left(\vec{h}^H\vec{v}_1^t\right)\rbrace$ and $\hat{\beta}_k^{(t+1)}=\hat{p}_k^t\vec{b}_k^H(\hat{\eta}^{(t+1)}\vec{v}_1^t+\vec{v}_2^t)$.
At each iteration of CGLS,  the main computational step  is a matrix-vector product $\vec{B}\vec{x}$ (here $\vec{x}$ is an arbitrary vector with length $M$). For the FMCW range-Doppler imaging problem, the matrix-vector product  $\vec{B}\vec{x}$  can be  efficiently calculated by  applying 2-D inverse FFT to $\vec{x}$.  
\subsubsection{Fast implementations of 1bLIKES and 1bIAA}  
The  computation of the covariance matrix $\vec{R}$ and its inverse cannot be avoided in 1bLIKES and 1bIAA. For the PMCW radar, the 2-D covariance matrix  $\vec{R}$ can be written as:
\begin{align}
    \vec{R}&=(\vec{\Phi}\otimes \vec{S})\vec{P}(\vec{\Phi}\otimes \vec{S})^H \nonumber \\
    &=\sum_{k_{\rm r}=1}^{K_{\rm r}}\sum_{k_{\rm d}=1}^{K_{\rm d}}p_{k_{\rm r},k_{\rm d}}\left[\vec{\phi}_{K_{\rm d}}(k_{\rm d})\vec{\phi}^H_{K_{\rm d}}(k_{\rm d})\right]\otimes \left[\vec{s}_{k_{\rm r}}\vec{s}_{k_{\rm r}}^H\right], \label{R:PMCW}
\end{align}
where $p_{k_{\rm r},k_{\rm d}}$ is the power at the range-Doppler grid point $(k_{\rm r},k_{\rm d})$.  It follows from (\ref{R:PMCW}) that $\vec{R}$ has a block Toeplitz structure, with $K_{\rm d} \times K_{\rm d}$ blocks each with $K_{\rm r}\times K_{\rm r}$ elements:
\begin{equation}
    \vec{R}=\left[
    \begin{matrix}
       \vec{R}_1 & \vec{R}_2 &\dots & \vec{R}_{K_{\rm d}} \\
      \vec{R}_{-2} & \vec{R}_1 & \dots & \vec{R}_{K_{\rm d}-1} \\
      \vdots &  \vdots & \ddots &\vdots \\
     \vec{R}_{-K_{\rm d}} & \vec{R}_{-K_{\rm d}+1} & \dots & \vec{R}_1 
    \end{matrix}    
    \right].
\end{equation}
The Gohberg-Semencul  factorization can be used to efficiently compute $\vec{R}^{-1}$ via FFTs \cite{akaike1973block,stoica1997}. For the range-Doppler imaging via  LFMCW radar, the 2-D covariance matrix $\vec{R}$ has a  Toeplitz-block-Toeplitz  structure. The fast implementations of 1bLIKES and 1bIAA  for the  LFMCW radar   can   follow the steps provided in  \cite{xue2011iaa}. 
\section{Simulated and experimental Examples}
We now evaluate the performance of the proposed one-bit weighted SPICE algorithms via both simulated and experimental examples. We first focus on range estimation using an LFMCW radar  and then shift our attention to range-Doppler imaging via  a PMCW radar. Finally, we present an experimental example of using one-bit weighted SPICE for range-Doppler imaging using measured LFMCW radar data. All  examples were run on a PC with  Intel(R)  i5-8250U @1.60 GHz CPU and 8.0 GB RAM. 
\subsection{Simulated Examples}
\subsubsection{Range  estimation via LFMCW radar}
\begin{figure*}[htbp]
\centering
\subfigure[1bPER]{
\label{fig:sin_mf}
\begin{minipage}[t]{0.32\linewidth}
\centering
\centerline{\epsfig{figure=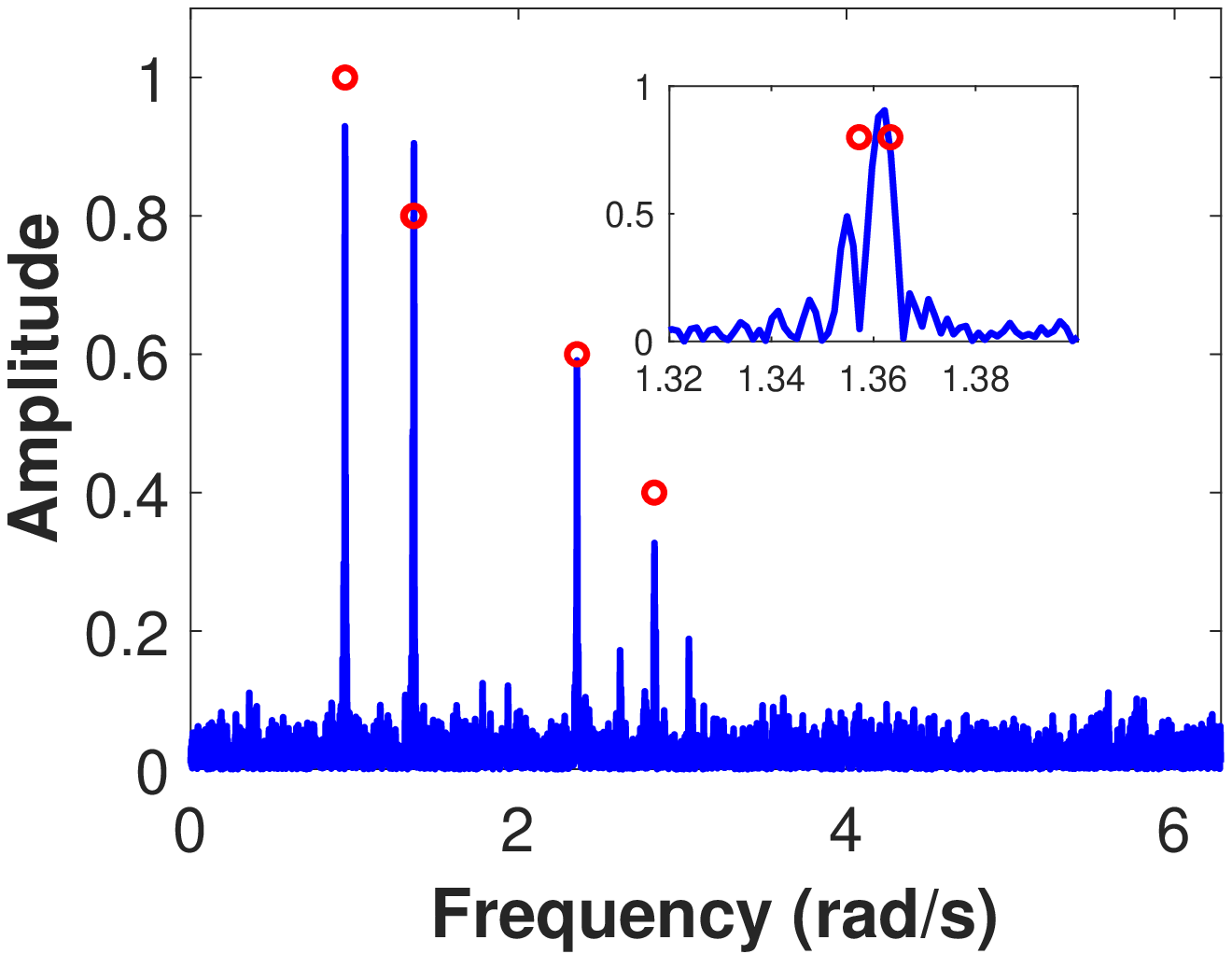,width=5.5cm}}
\end{minipage}}
\subfigure[ADMM ]{
\label{fig:sin_admm}
\begin{minipage}[t]{0.32\linewidth}
\centering
\centerline{\epsfig{figure=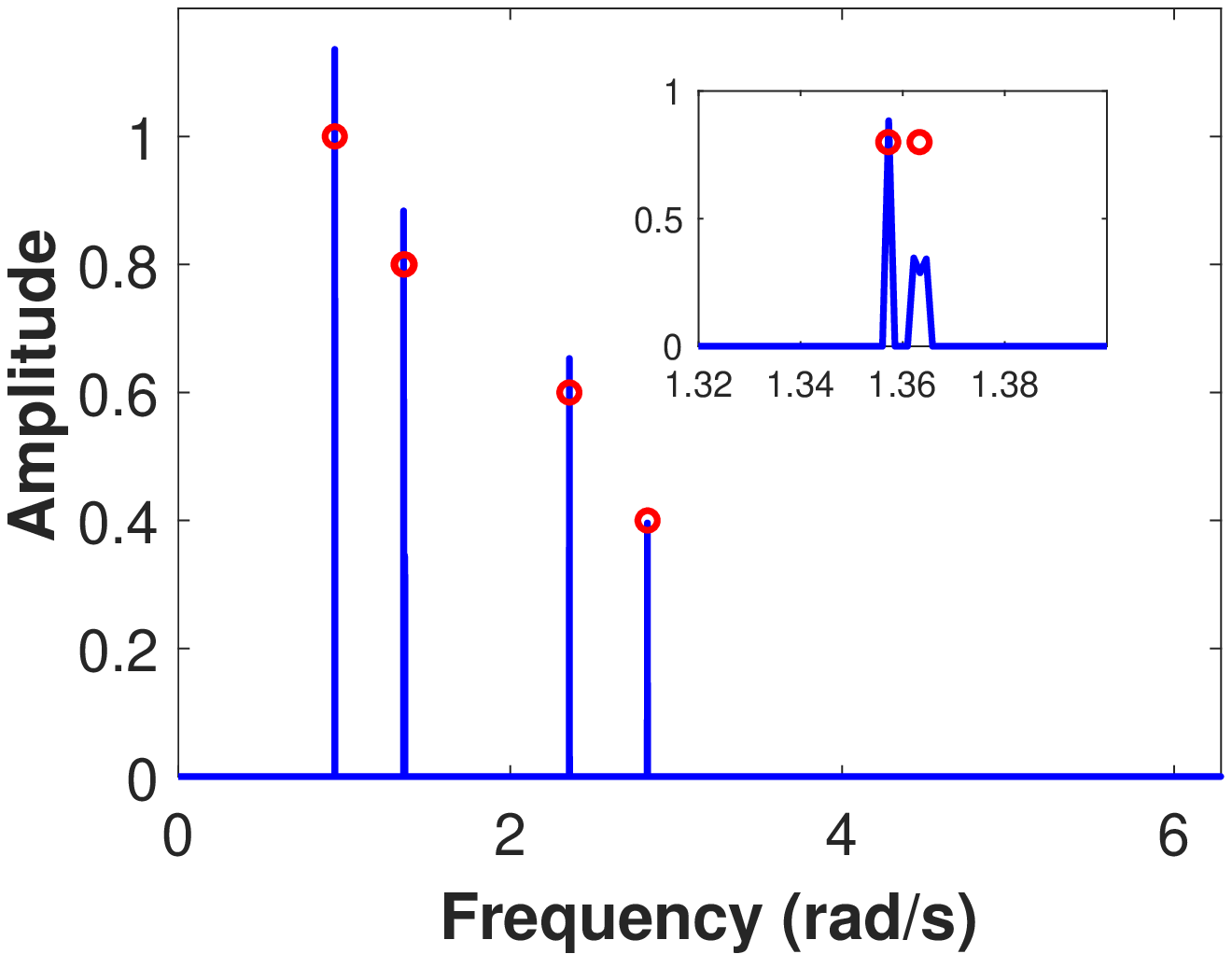,width=5.5cm}}
\end{minipage}}
\subfigure[1bSPICE]{
\label{fig:sin_spice}
\begin{minipage}[t]{0.32\linewidth}
\centering
\centerline{\epsfig{figure=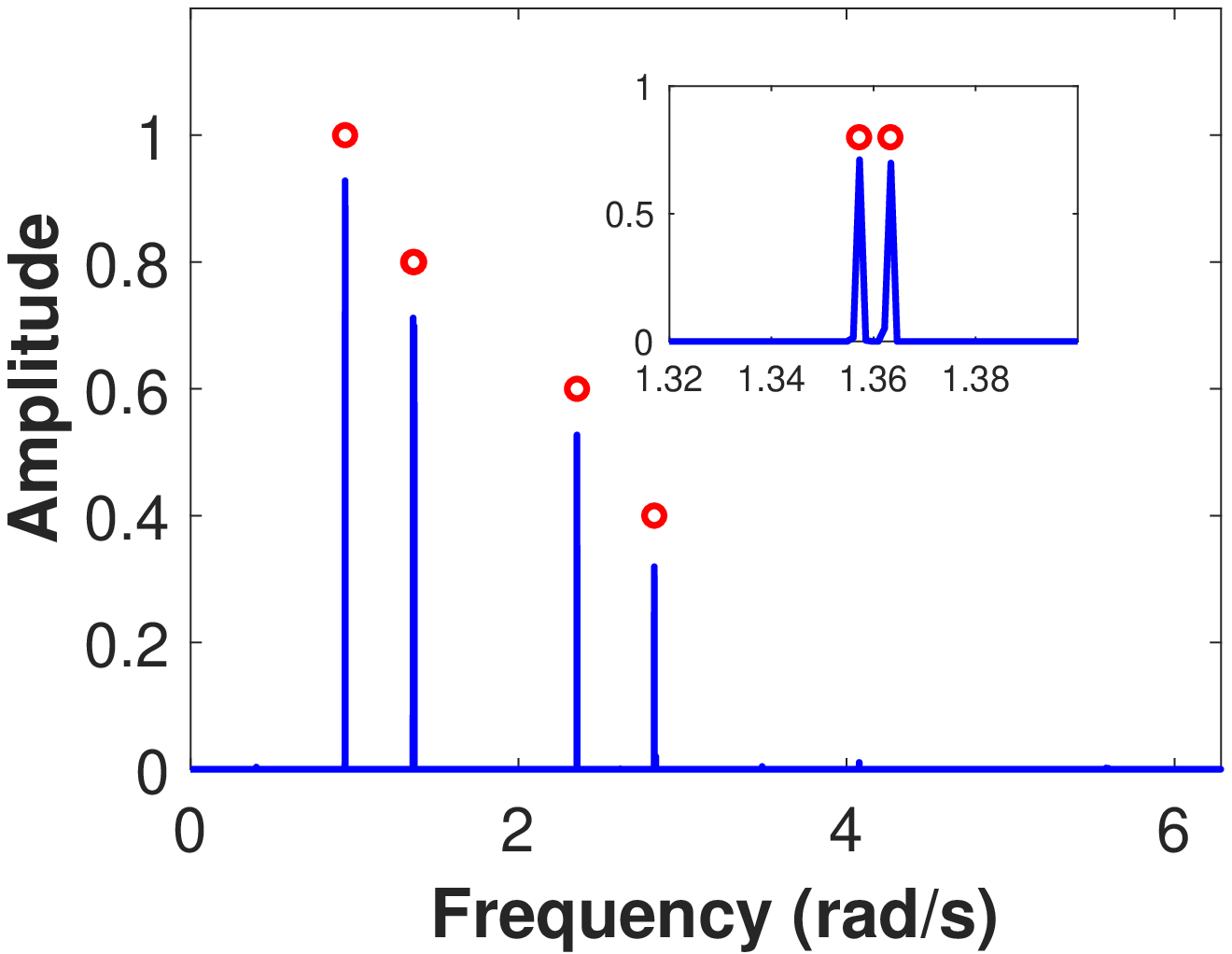,width=5.5cm}}
\end{minipage}}
\subfigure[1bLIKES]{
\label{fig:sin_likes}
\begin{minipage}[t]{0.32\linewidth}
\centering
\centerline{\epsfig{figure=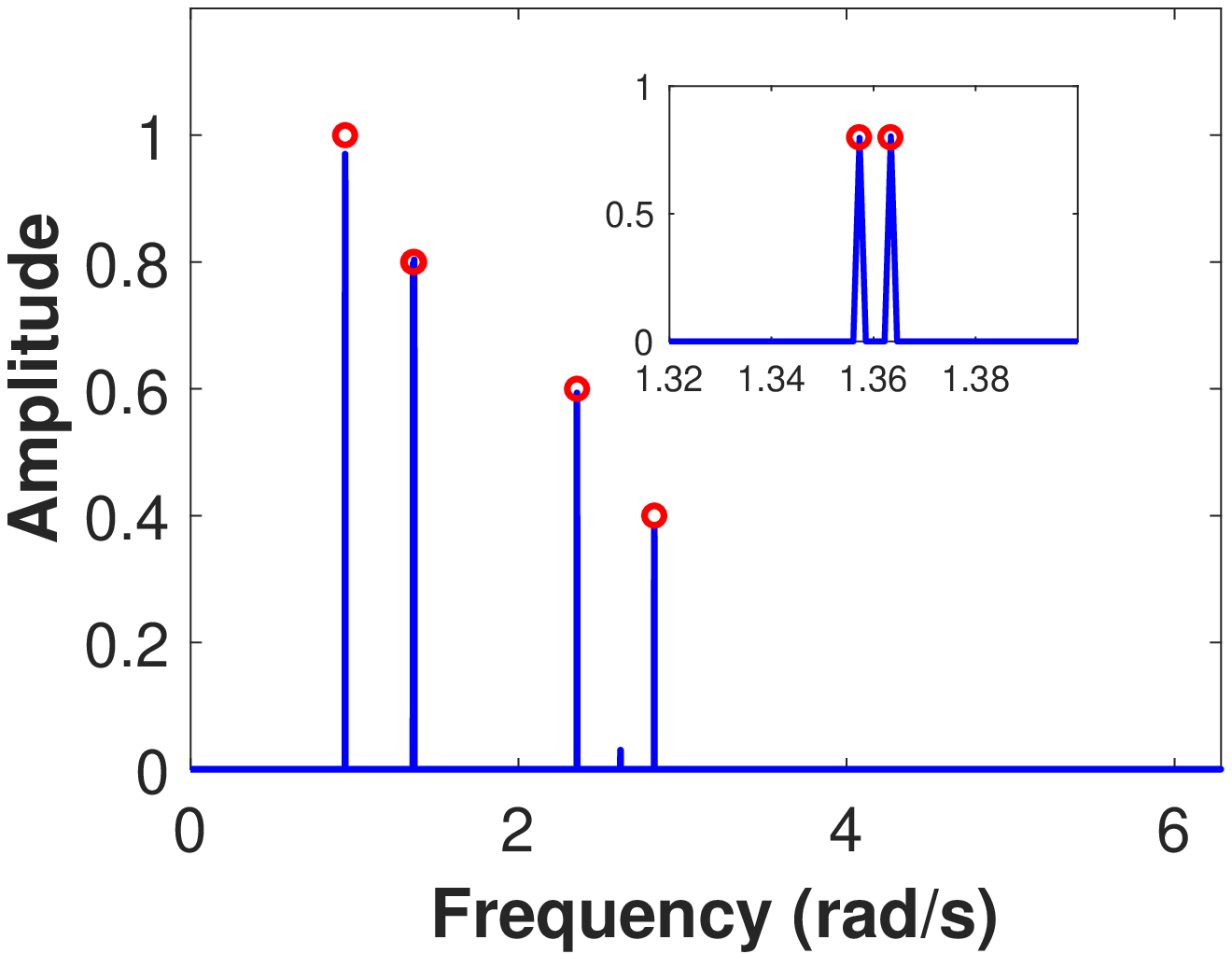,width=5.5cm}}
\end{minipage}}
\subfigure[1bSLIM]{
\label{fig:sin_slim}
\begin{minipage}[t]{0.32\linewidth}
\centering
\centerline{\epsfig{figure=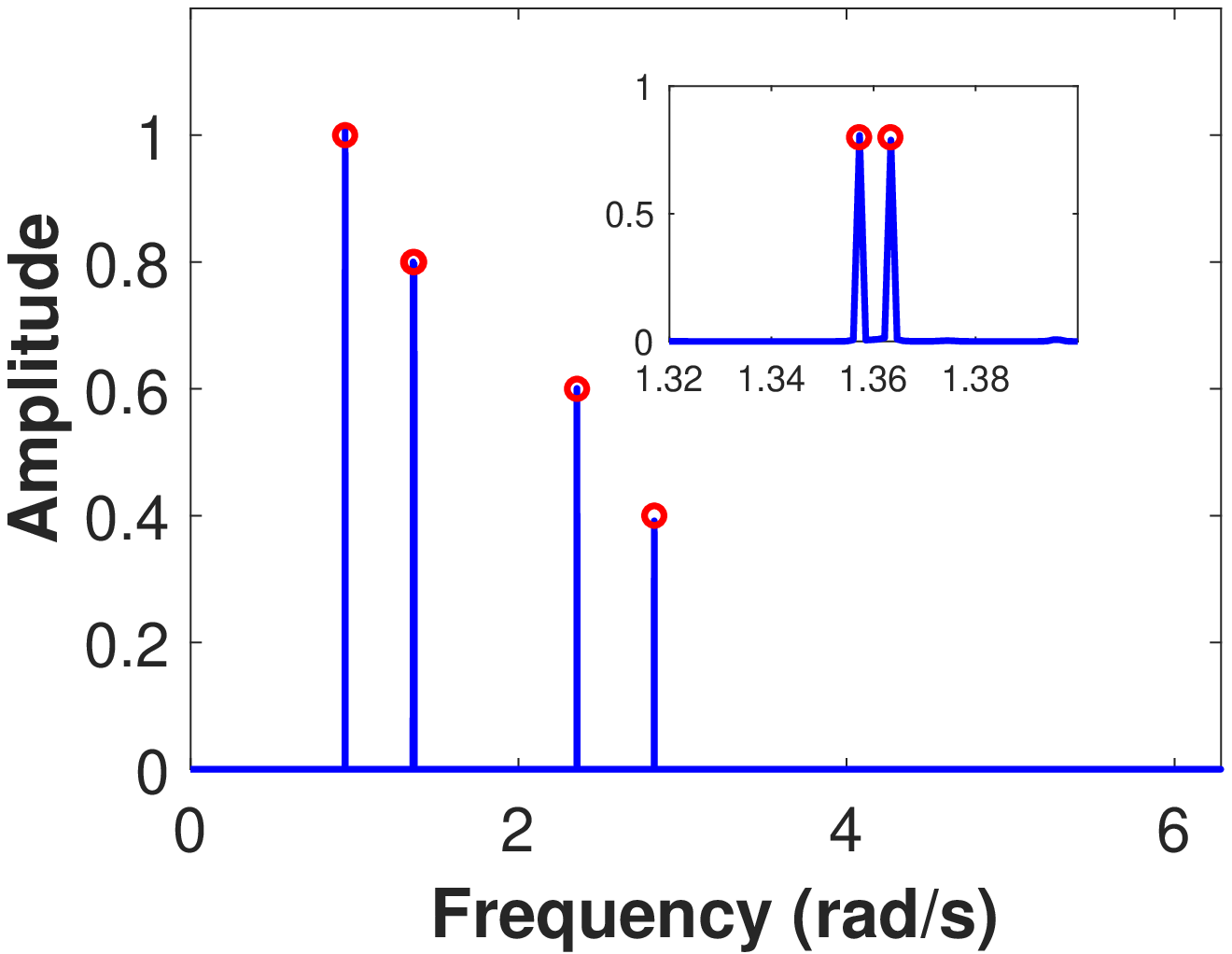,width=5.5cm}}
\end{minipage}}
\subfigure[1bIAA]{
\label{fig:sin_iaa}
\begin{minipage}[t]{0.32\linewidth}
\centering
\centerline{\epsfig{figure=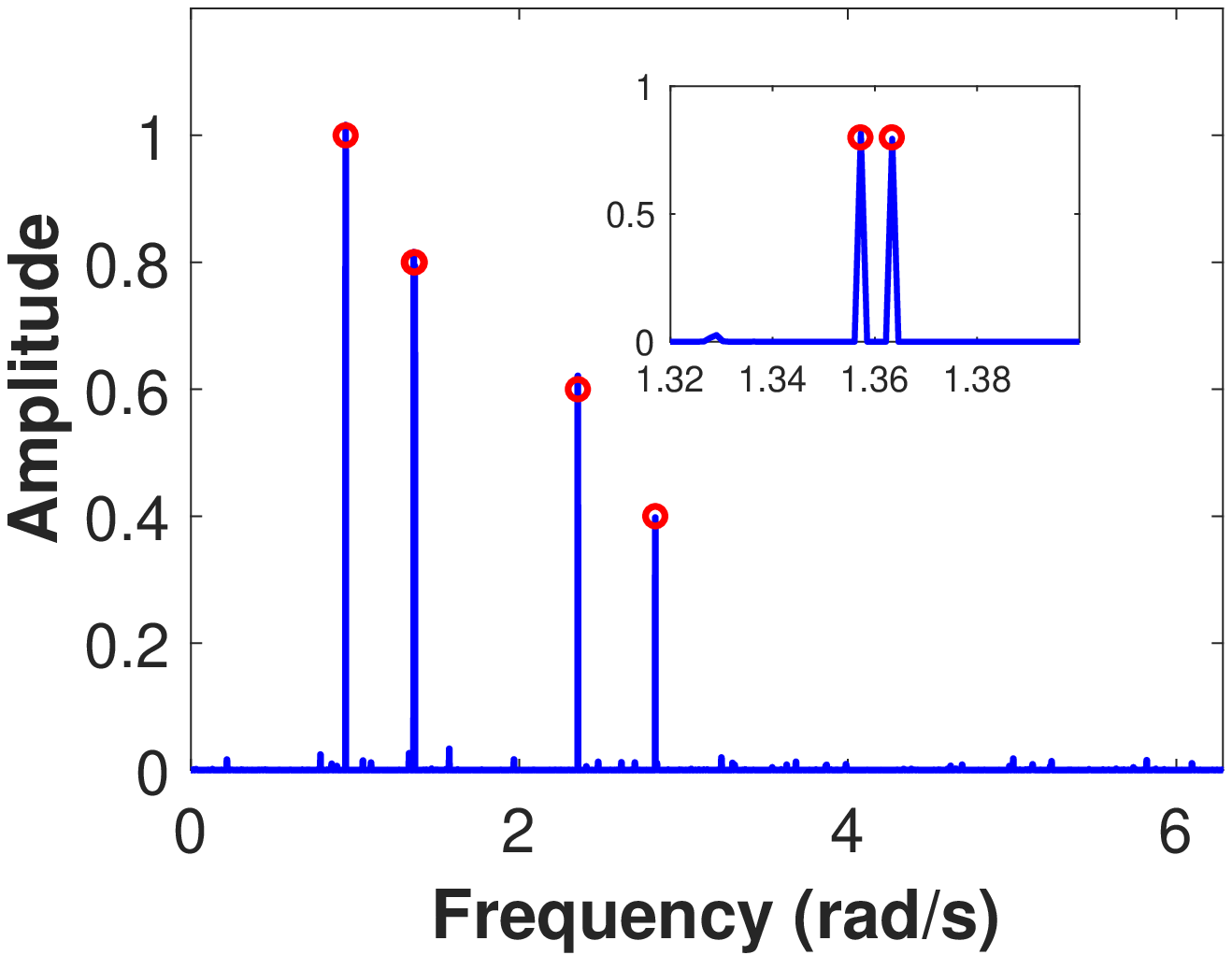,width=5.5cm}}
\end{minipage}}
\caption{Spectra obtained via (a) 1bPER, (b) ADMM, (c) 1bSPICE, (d) 1bLIKES, (e) 1bSLIM, (f) 1bIAA. The circles in the plots indicate the true values of the amplitudes.}
\label{fig:sin}
\end{figure*}

As introduced in Section \ref{section:LFMCW}, the range  estimation problem using an LFMCW radar corresponds to a 1-D  sinusoidal  parameter  estimation problem. The length $N=1024$ signal we  consider consists of five sinusoids with  amplitudes  $\lbrace \alpha_k \rbrace_{k=1}^5=\lbrace 1, 0.8, 0.8, 0.6, 0.4 \rbrace$, frequencies 
$\lbrace 0.150 \times 2\pi, 0.216 \times 2\pi, (0.216+1/ N)\times 2\pi, 0.375\times 2\pi, 0.450 \times 2\pi \rbrace$, and phases selected independently and randomly from a uniform distribution on $\left[0,2\pi\right]$.  The number of grid points we use in the frequency domain is $M=5N$.  Note that  the true  frequencies of the second and third sinusoids are not exactly on the frequency grid, and their  frequency separation is equal to the frequency resolution limit of the conventional periodogram. The time-varying threshold has the  real and imaginary parts selected randomly and equally likely from a predefined eight-element set $\lbrace -h_{\max}, -h_{\max}+\Delta, \dots,h_{\max}-\Delta,h_{\max} \rbrace$ with $h_{\max}=\frac{\sqrt{\sum_{k=1}^5\alpha_k^2+\sigma^2}}{2}$ and $\Delta=\frac{2h_{\max}}{7}$. Note that in practice, a rough estimate of the received signal  power (i.e., $\sum_{k=1}^5\alpha_k^2+\sigma^2)$ can be obtained from an analog AGC circuit \cite{mo2017channel}. The threshold is used to obtain the signed measurements. The SNR, which is defined as $10\log_{10}\frac{\sum_{k=1}^5 \alpha_k^2}{\sigma^2}$,  is set to 20 dB. \\
\indent In Fig. \ref{fig:sin_mf}, we display the amplitude of the  spectral estimate obtained via the one-bit periodogram \cite{gianelli2017one}, referred to as 1bPER.  1bPER suffers from significant  leakage (i.e., high sidelobes)  and poor resolution problems (it cannot resolve the second and  third sinusoids). Fig. \ref{fig:sin_admm} shows the result obtained using  the ADMM log-norm algorithm \cite{heng}, referred to as ADMM.  The ADMM amplitude estimates  are usually overestimated and this algorithm also has a resolution problem similar to that of 1bPER. The one-bit weighted SPICE results in Figs. \ref{fig:sin_spice}--\ref{fig:sin_iaa} show improved resolutions and reduced sidelobes. We observe that 1bSPICE has a slight amplitude underestimation problem,  while 1bLIKES, 1bSLIM and 1bIAA provide accurate amplitude estimates. Similar to the behavior of the original weighted SPICE algorithms,  the results of 1bLIKES and 1bSLIM are more sparse than that of 1bIAA.   \\
\begin{figure}[htp]
\subfigure[]{
\label{fig:mse1}
\begin{minipage}[t]{0.48\linewidth}
\centerline{\epsfig{figure=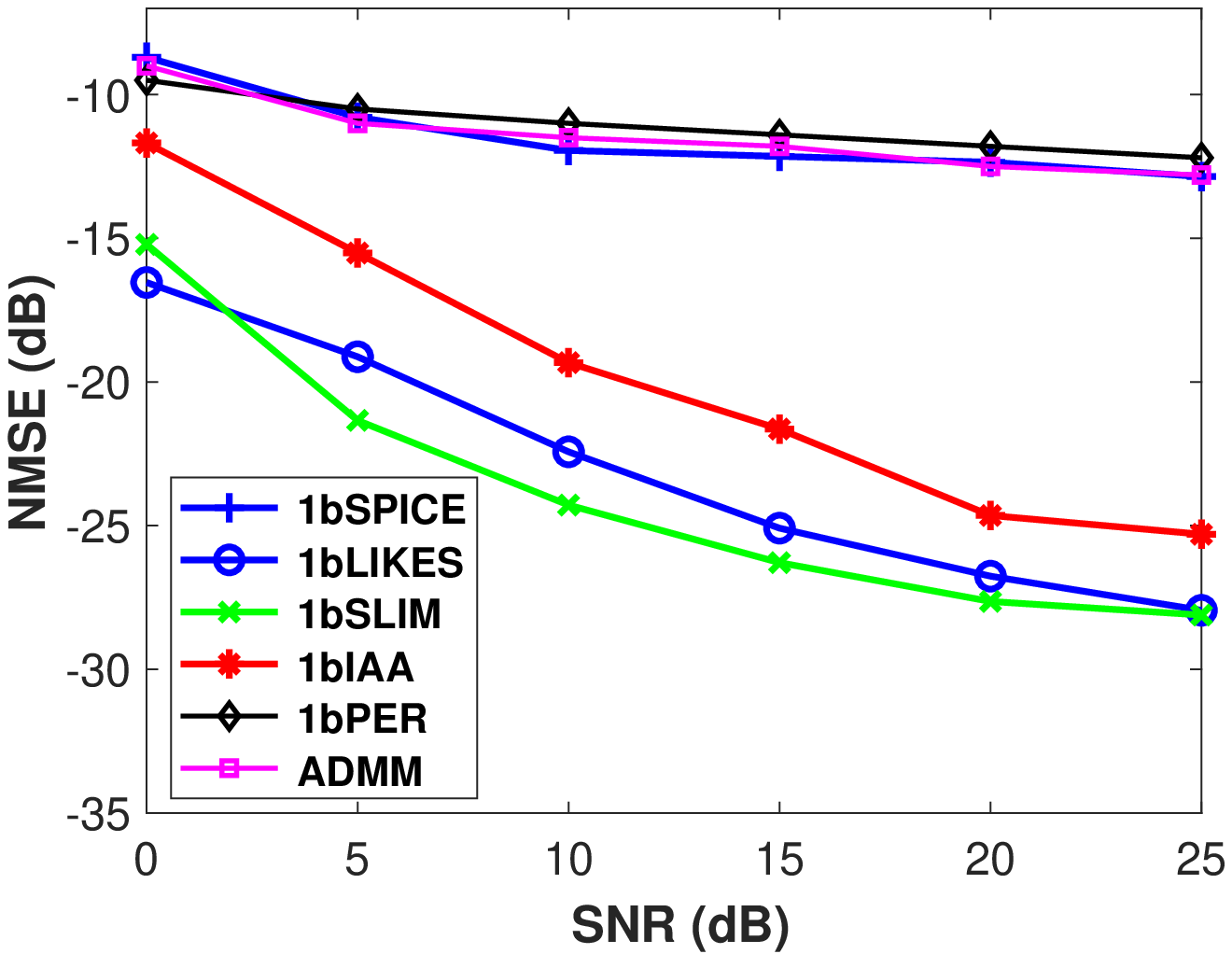,width=4.9cm}}
\end{minipage}}
\subfigure[]{
\label{fig:sidelobe}
\begin{minipage}[t]{0.48\linewidth}
\centerline{\epsfig{figure=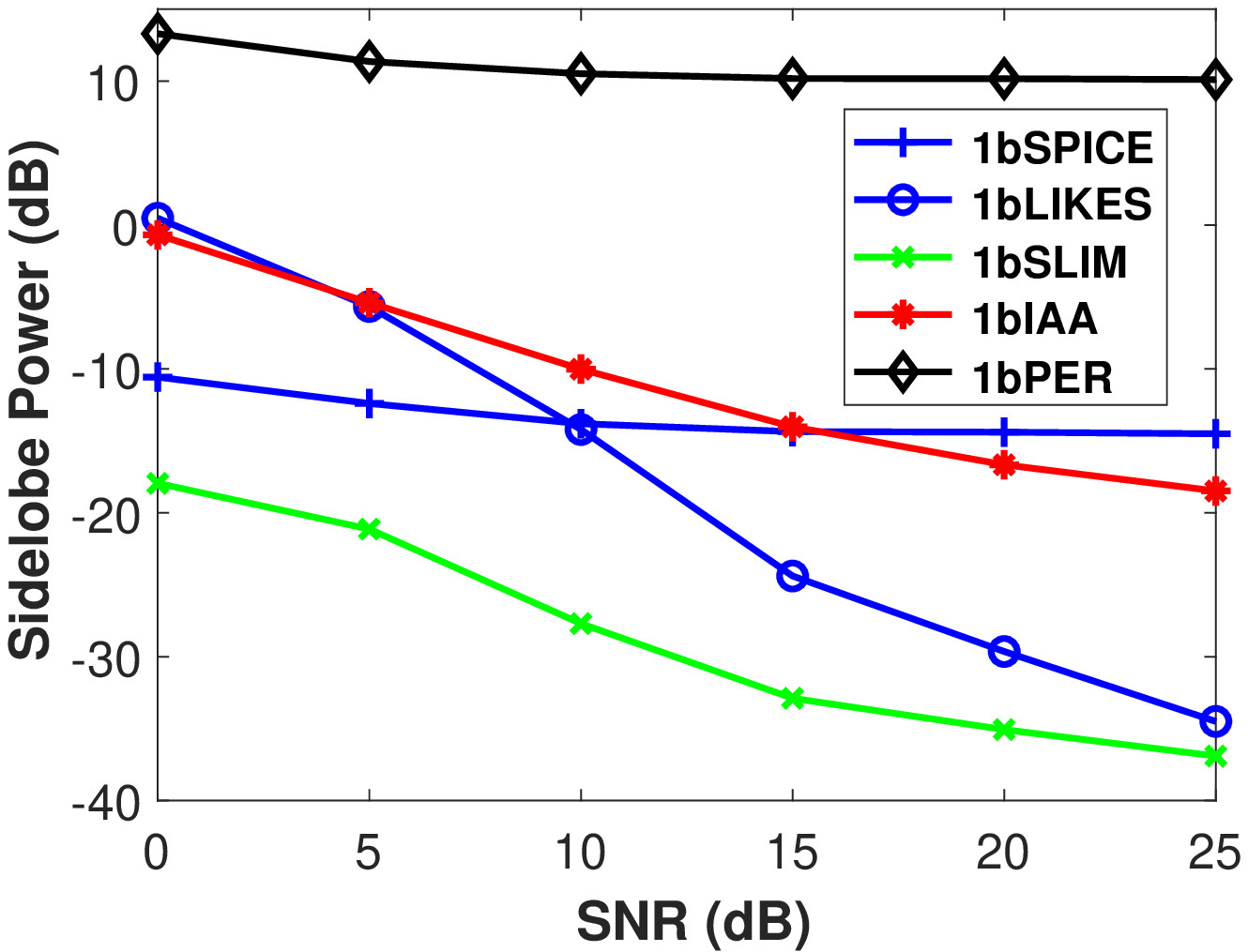,width=4.9cm}}
\end{minipage}}
\caption{Estimation performance of 1bPER, ADMM, 1bSPICE, 1bLIKES, 1bSLIM and 1bIAA (from 500 Monte Carlo runs): (a) average NMSEs versus SNR, (b) average sidelobe power versus SNR.  }
\label{fig:mse}
\end{figure}
\indent To further illustrate the performance of the algorithms, we show the average  NMSEs of the amplitude estimates  as well as the average sidelobe powers, obtained  from 500 Monte Carlo runs,  as functions of SNR in Fig. \ref{fig:mse1} and Fig. \ref{fig:sidelobe}.  The average NMSE and sidelobe power $P_{\rm s}$  are calculated as: 
\begin{align}
    &{\rm NMSE}=\frac{1}{2500}\sum_{k=1}^5\sum_{i=1}^{500}\frac{ |\hat{\alpha}_k^{(i)} -\alpha_k|^2}{\alpha_k^2}, \nonumber \\
    & P_{\rm s}=\frac{1}{500}\sum_{i=1}^{500} \left(\Vert \hat{\vec{\gamma}}^{(i)} \Vert^2-\sum_{k=1}^5\left( \hat{\alpha}^{(i)}_k \right)^2 \right), \nonumber
\end{align}
where $\hat{\alpha}_k^{(i)}$ denotes the estimate of $\alpha_k$ in the $i$th Monte Carlo run; $\hat{\vec{\gamma}}^{(i)}$ represents the estimated sparse parameter vector in the $i$th Monte Carlo run and each Monte Carlo run corresponds to an  independent noise and threshold realization.
Inspecting the results,  we see that 1bPER  gives  poor  amplitude estimates and high sidelobe levels. Furthermore, as the SNR increases, there is little improvement in its amplitude estimation accuracy and sidelobe levels. The ADMM  approach provides slightly  better results than 1bPER. Since ADMM provides a very  sparse spectrum, we do not plot its sidelobe power. 1bSPICE provides competitive sidelobe levels, but its  amplitude estimates are biased downward. 1bLIKES and 1bSLIM have similar amplitude estimation performances, and 1bSLIM has lower sidelobe levels at low SNRs. In this example, 1bIAA  performs slightly worse than 1bSLIM and 1bLIKES in terms of  both amplitude estimation accuracy  and sidelobe levels.
\begin{figure}[htp]
\centering
\includegraphics[width=0.35\textwidth]{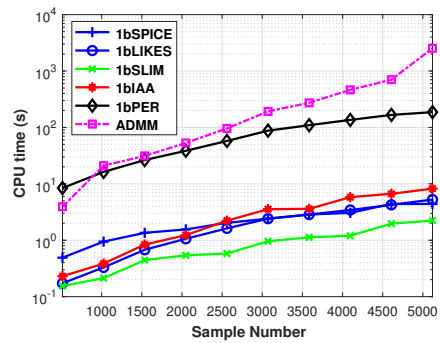}
\caption{Average computation time versus the sample number for 1bPER, ADMM, 1bSPICE, 1bLIKES, 1bSLIM and 1bIAA for SNR=20 dB.}
\label{fig:time}
\end{figure}
\par The average computation times  needed by the aforementioned algorithms are  plotted on a logarithmic scale in Fig. \ref{fig:time}  versus the sample number $N$  and for a number of grid points $M$ equal to $5N$.  The fast implementations of 1bSPICE and 1bSLIM use the CGLS approach  with FFT. The fast implementations of 1bLIKES and 1bIAA  make use of the Gohberg–Semencul factorization as in \cite{xue2011iaa}.  Note that all  four proposed algorithms are more than an order of magnitude  faster than  the  1bPER and ADMM  algorithms.  Among the  proposed algorithms, 1bSLIM is  faster than the other three algorithms,  which is due to the lower computational complexity of each iteration of 1bSLIM  as well as its faster convergence speed.  1bSLIM can converge within 30 iterations, whereas the other algorithms tend to require more iterations.  1bSPICE and 1bLIKES  have similar computation times, while 1bIAA tends to be slower than the other algorithms as the sample number $N$ increases. 
\begin{figure*}[htb]
\centering
\subfigure[1bPER]{
\label{fig:pmcw_1bperiodogram}
\begin{minipage}[t]{0.30\linewidth}
\centering
\centerline{\epsfig{figure=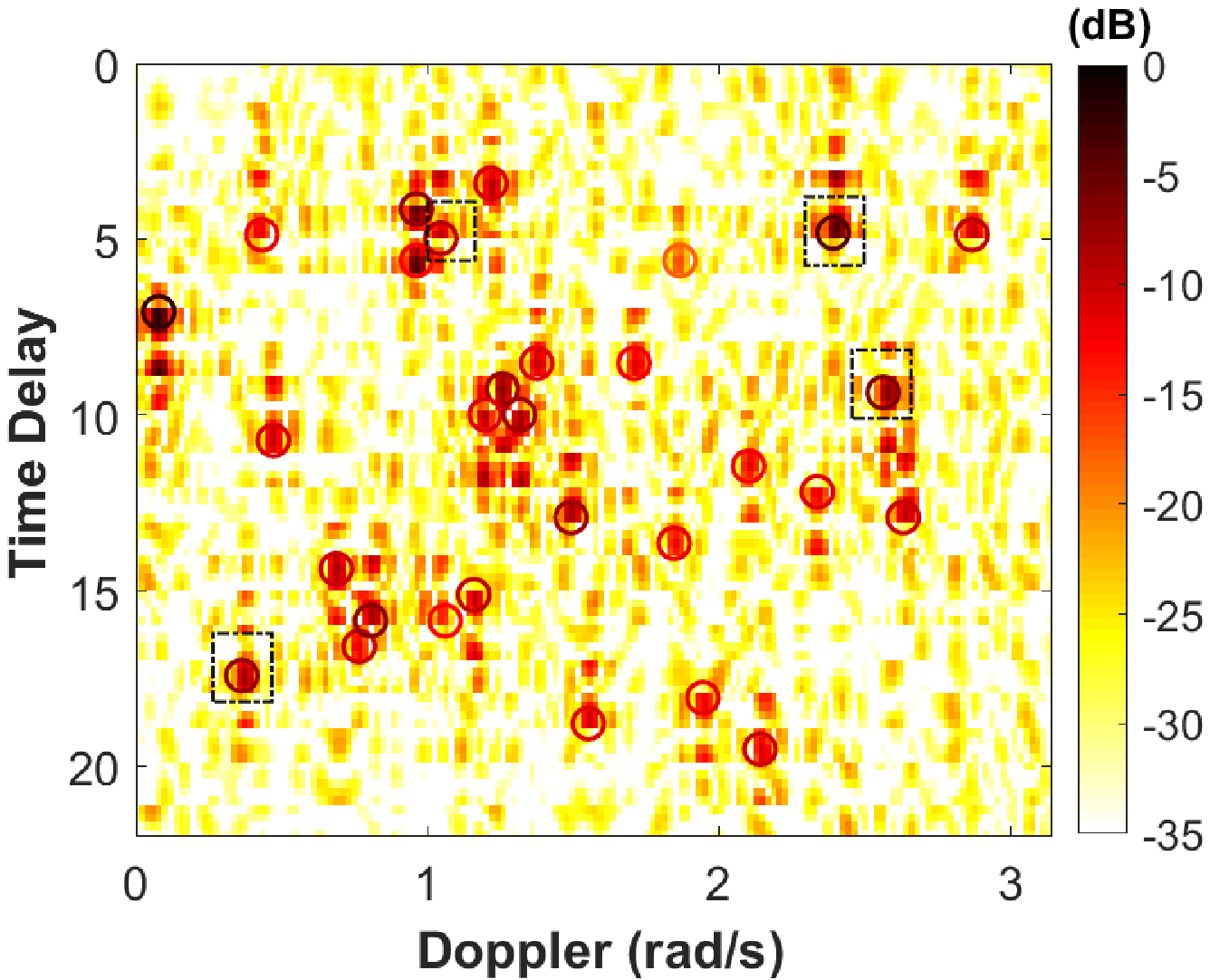,width=5.5cm}}
\end{minipage}}
\subfigure[ADMM]{
\label{fig:pmcw_admm1}
\begin{minipage}[t]{0.30\linewidth}
\centering
\centerline{\epsfig{figure=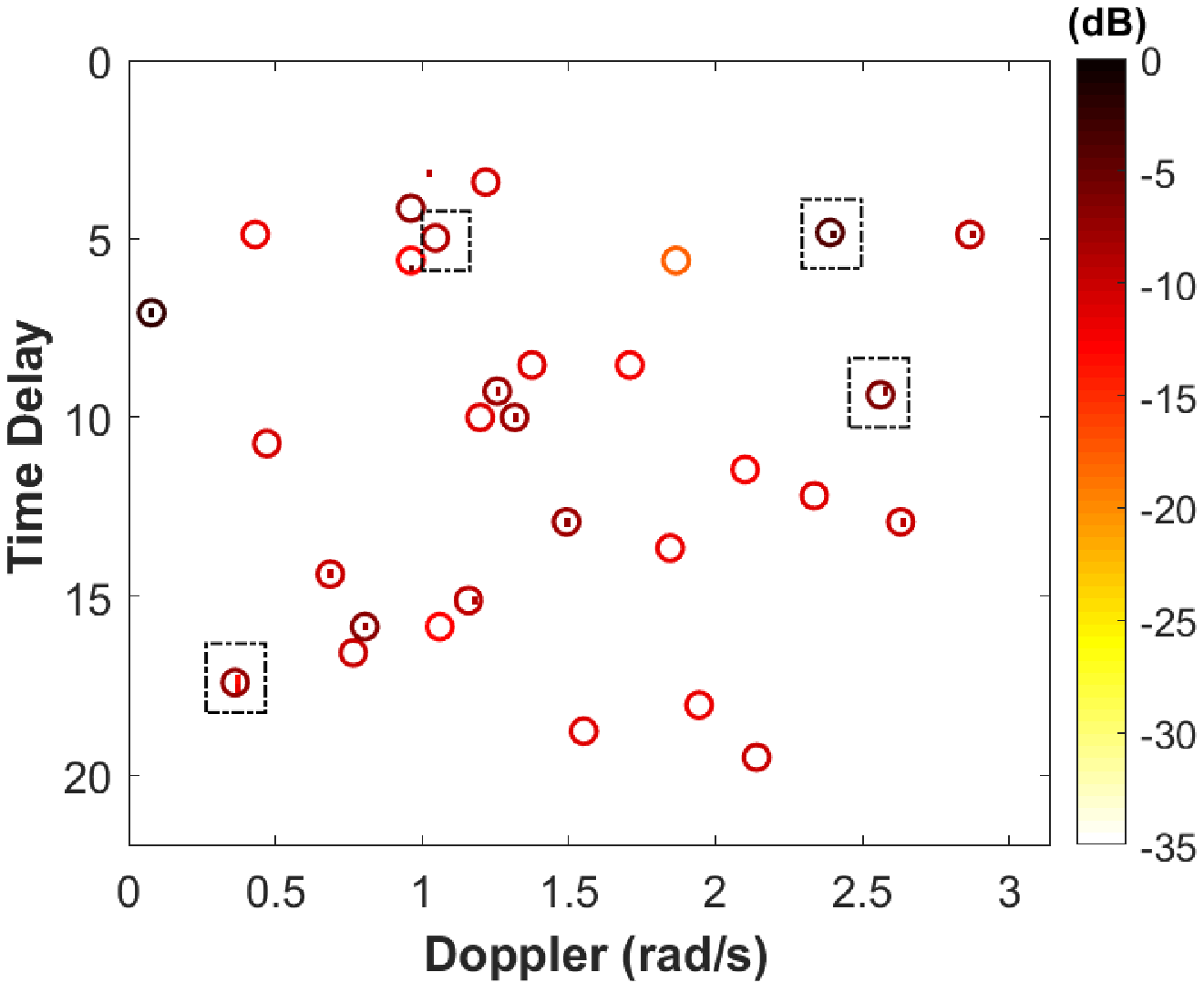,width=5.5cm}}
\end{minipage}}
\subfigure[1bSPICE]{
\label{fig:pmcw_spice}
\begin{minipage}[t]{0.30\linewidth}
\centering
\centerline{\epsfig{figure=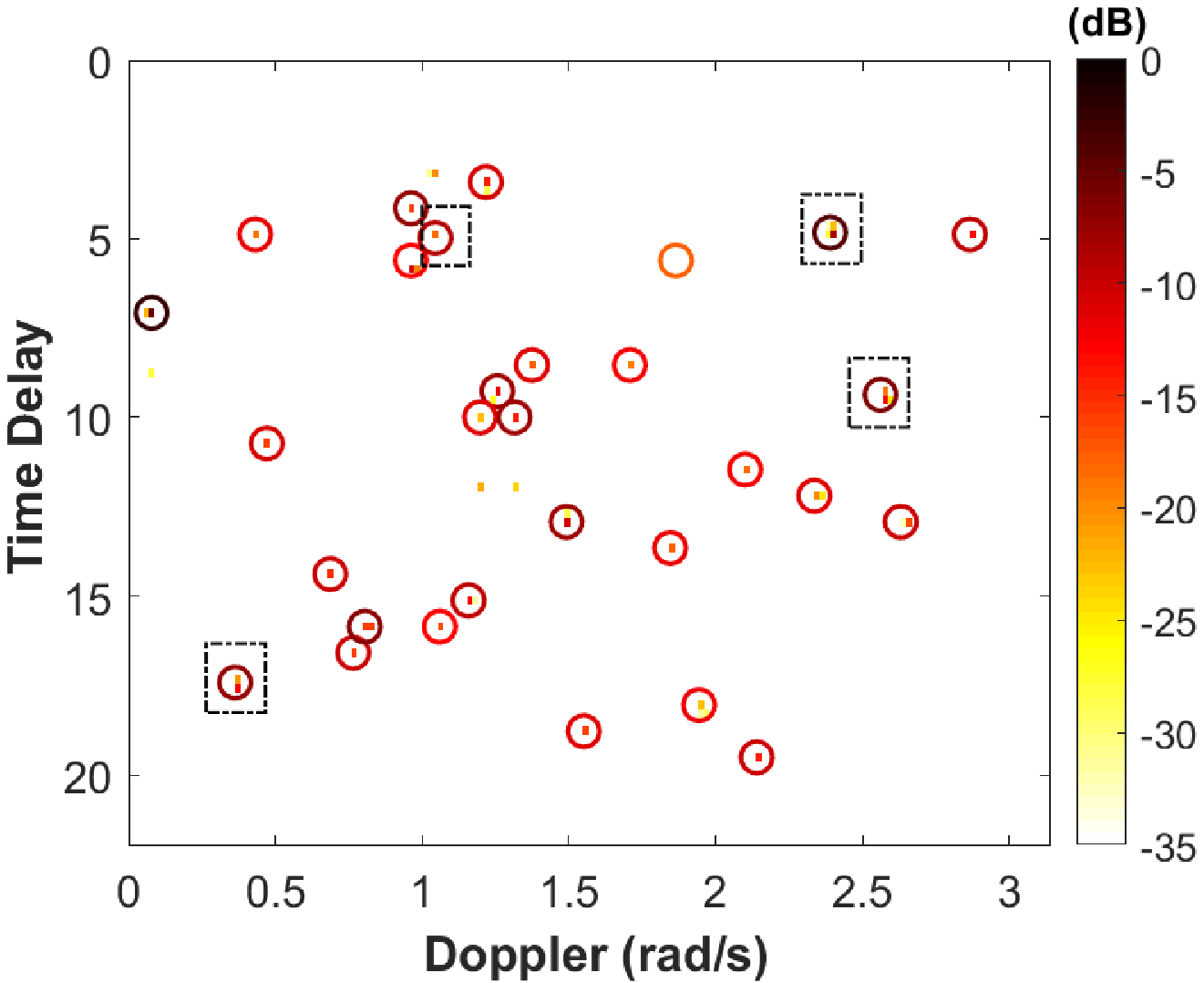,width=5.5cm}}
\end{minipage}}
\subfigure[1bLIKES]{
\label{fig:pmcw_likes}
\begin{minipage}[t]{0.30\linewidth}
\centering
\centerline{\epsfig{figure=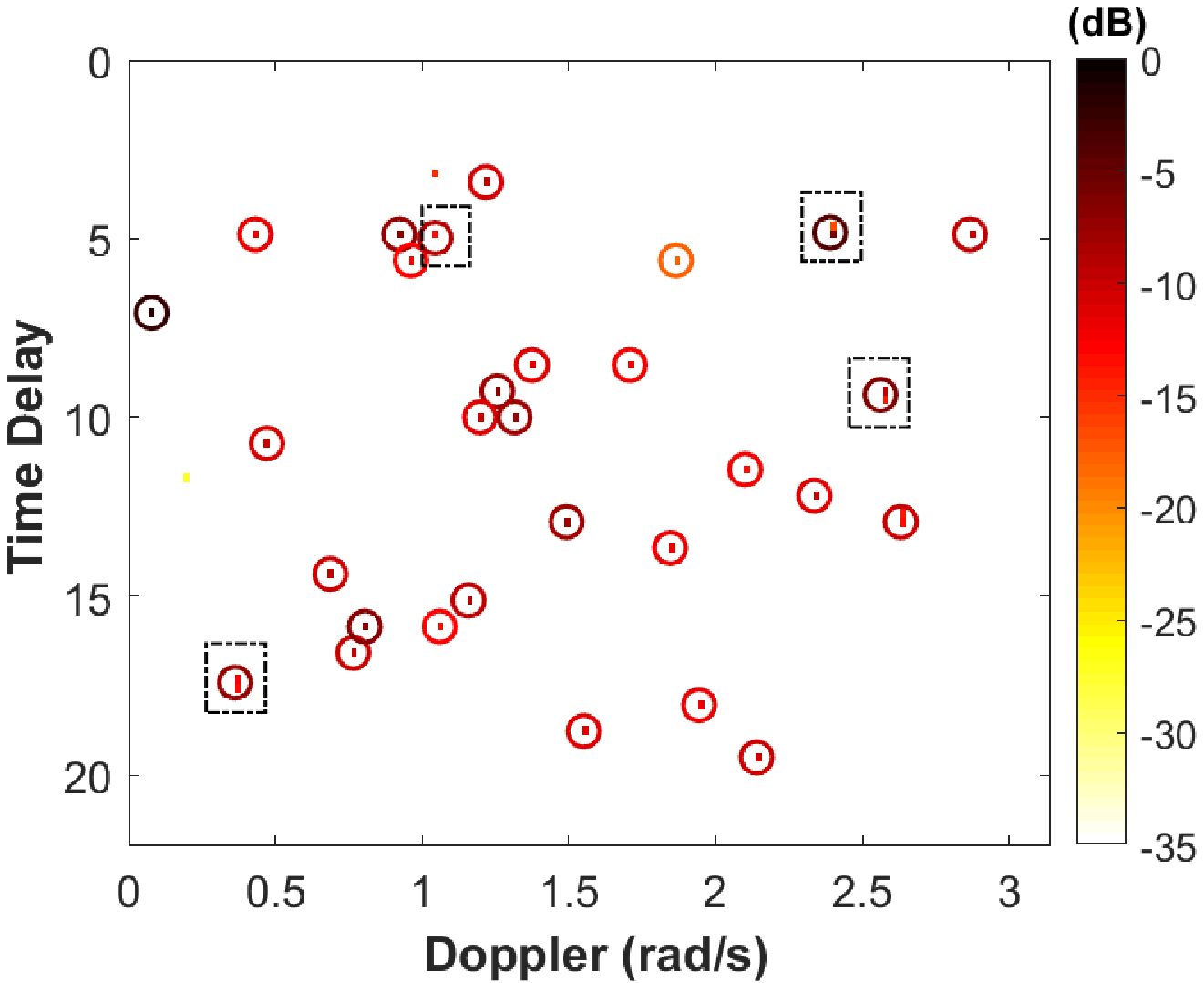,width=5.5cm}}
\end{minipage}}
\subfigure[1bSLIM]{
\label{fig:pmcw_slim}
\begin{minipage}[t]{0.30\linewidth}
\centering
\centerline{\epsfig{figure=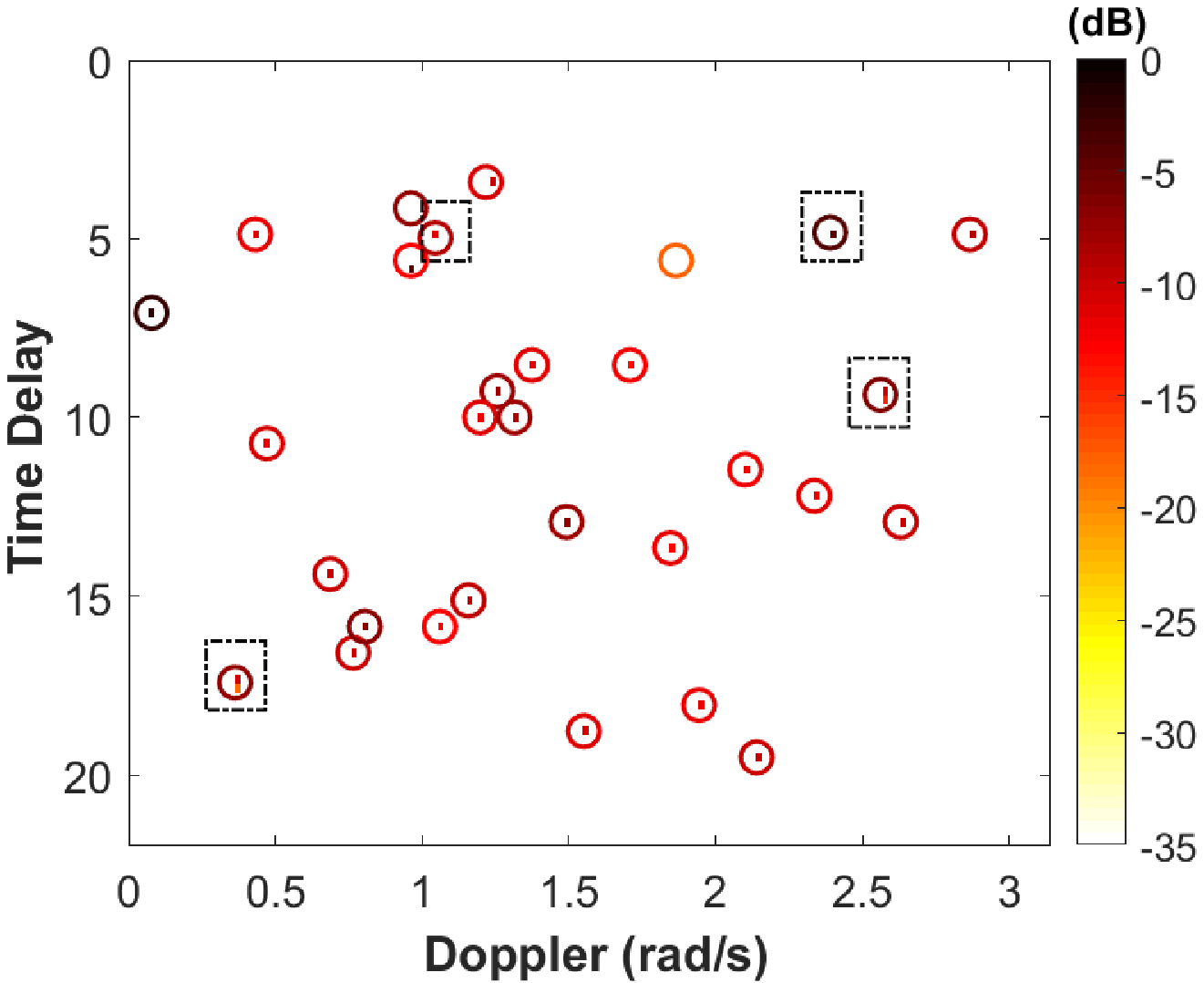,width=5.5cm}}
\end{minipage}}
\subfigure[1bIAA]{
\label{fig:pmcw_iaa}
\begin{minipage}[t]{0.30\linewidth}
\centering
\centerline{\epsfig{figure=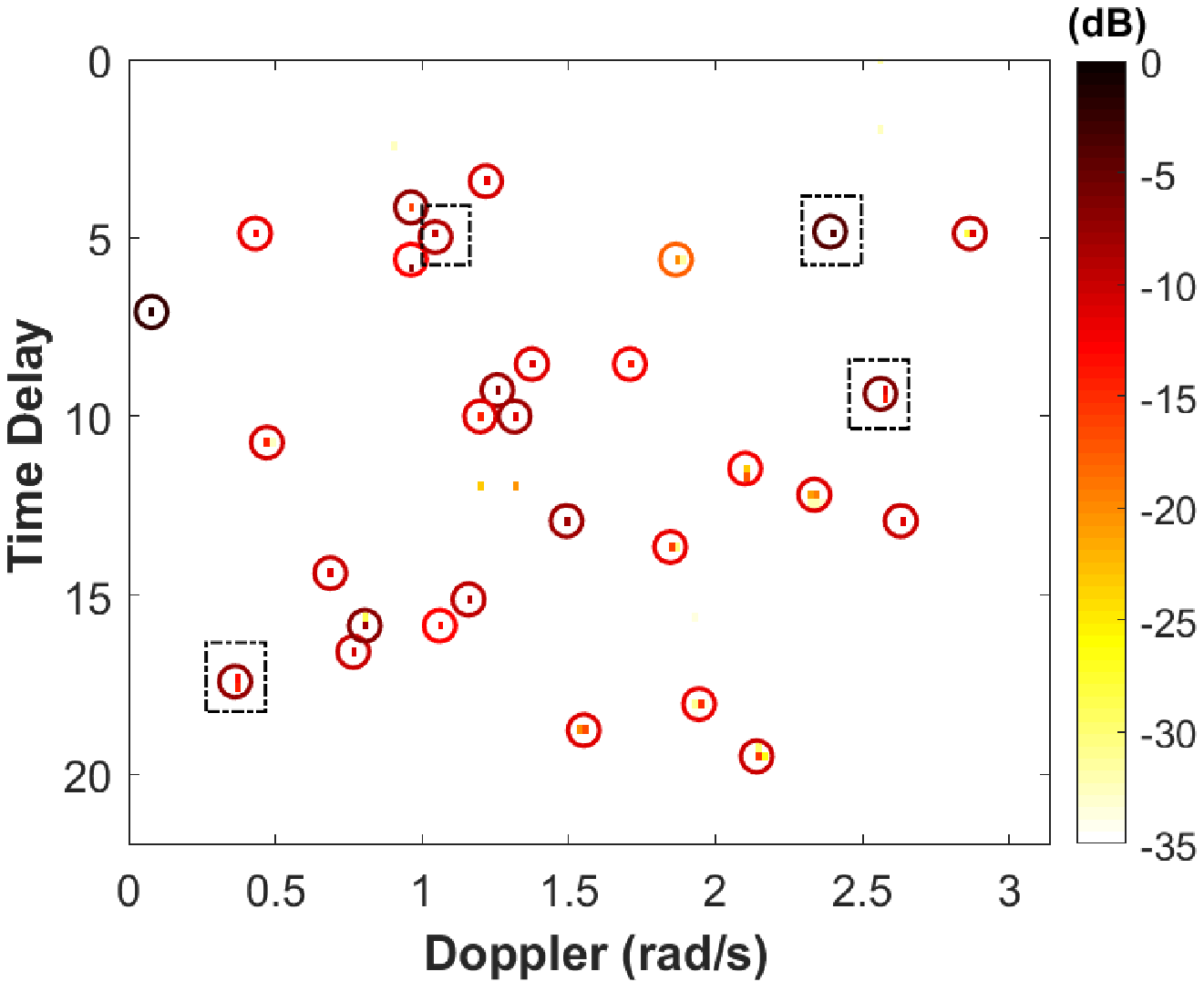,width=5.5cm}}
\end{minipage}}
\caption{The range-Doppler images of the one-bit PMCW radar for $N_1=31$, $N_2=64$, $K_r=4N_1$, $K_d=5N_2$, and SNR=15 dB.  (a) 1bPER, (b) ADMM, (c) 1bSPICE, (d) 1bLIKES, (e) 1bSLIM, (f) 1bIAA.  ``$\bigcirc$'' indicates the locations of the true targets (color-coded according to power, in dB). The dash-dot rectangles indicate the off-grid targets.}
\label{fig:pmcw}
\end{figure*}
\subsubsection{Range-Doppler imaging for PMCW radar}
\indent  For the PMCW radar, we use a maximum length sequence  (m-sequence) with length $N_1=31$ and $N_2=64$ periods within a CPI as the probing waveform. The number of grid points in the range and Doppler domains are set to $K_{\rm r}=4N_1$, $K_{\rm d}=5N_2$, respectively. The scene of interest consists of 30 targets, with their range-Doppler locations and powers indicated by the color-coded `$\bigcirc$' in Fig. \ref{fig:pmcw}. Note that there are four off-grid targets marked with dash-dot rectangles.  The amplitudes of the targets $\lbrace \alpha_k\rbrace_{k=1}^{30}$ are selected randomly between 0.1 and 1. In practical applications, the  time-varying threshold can be generated by a low-rate and low-precision DAC. To reduce the sampling rate of the DAC for cost reasons, we use a PRI-varying  threshold, that is the threshold is constant within each PRI and only changes from one PRI to another. The real and imaginary parts of the PRI-varying threshold are selected  randomly and equally likely   from a predefined eight-element set $\lbrace-h_{\max}, -h_{\max}+\Delta, \dots, h_{\max}-\Delta, h_{\max} \rbrace$ with  $h_{\max}=\frac{\sqrt{\sum_{k=1}^{30}{\alpha_k^2}+\sigma^2}}{2}$ and $\Delta=\frac{2h_{\max}}{7}$. The SNR, which is defined as $10\log_{10}\frac{\sum_{k=1}^{30}\alpha_k^2}{\sigma^2}$, is set to 15 dB. 
\par The range-Doppler imaging result obtained by using  1bPER with the one-bit data is shown in Fig. \ref{fig:pmcw_1bperiodogram}. As expected, 1bPER  suffers from  high sidelobe problems. As a result, the weak targets are buried  in the sidelobes of strong targets. In Fig. \ref{fig:pmcw_admm1}, we show the result of  the  ADMM  approach \cite{heng}.  ADMM   misses quite a few weak targets. The range-Doppler image of 1bSPICE is shown   in Fig. \ref{fig:pmcw_spice}.  1bSPICE misses one weak target, but provides much lower sidelobe levels than 1bPER and more  accurate target estimates than ADMM. 
\begin{figure}[htp]
\centering
\subfigure[2-D FFT]{
\label{fig:fmcw_mf}
\begin{minipage}[t]{0.38\linewidth}
\centering
\centerline{\epsfig{figure=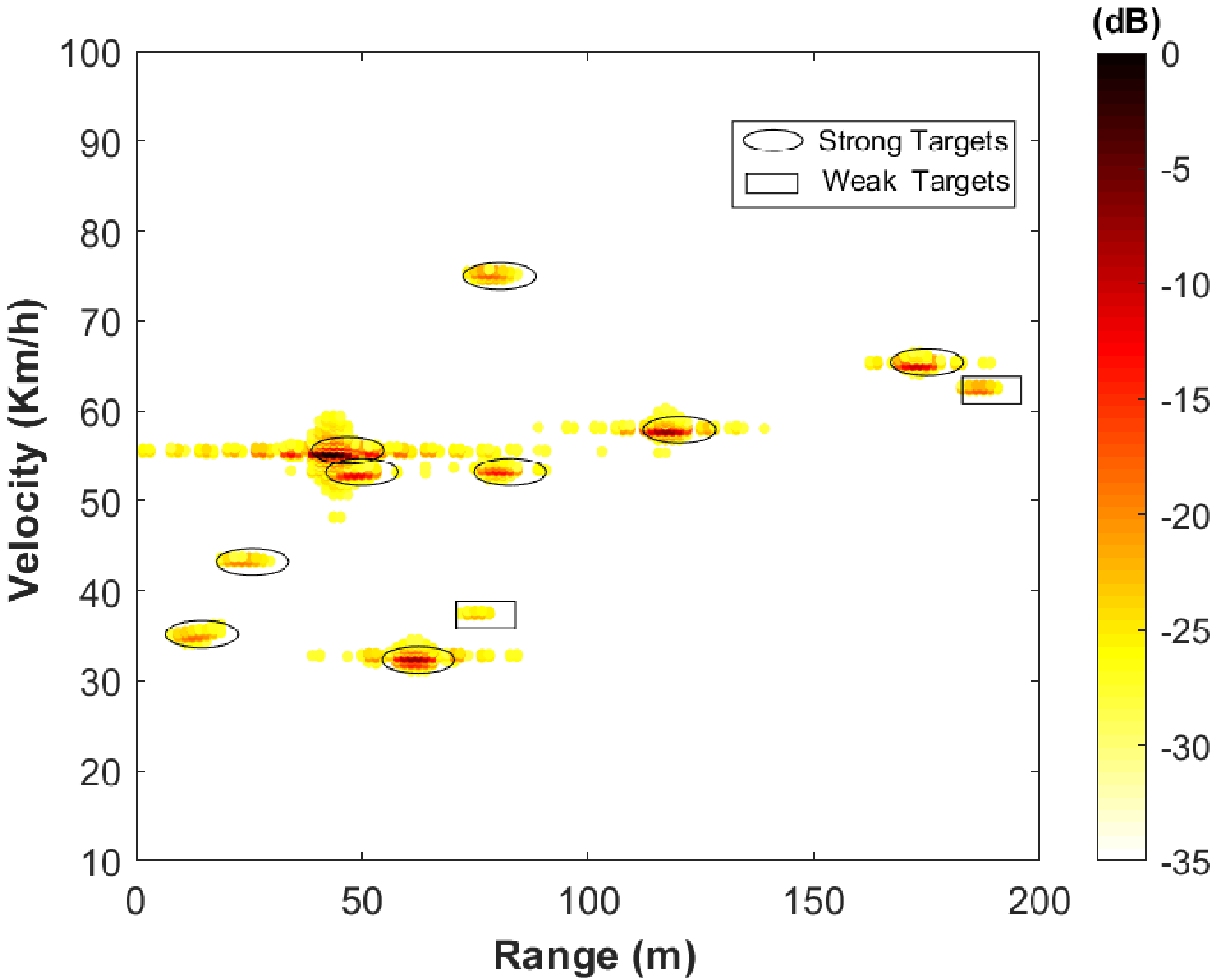,width=6.8cm}}
\end{minipage}} \\
\subfigure[The conventional IAA]{
\label{fig:fmcw_iaa}
\begin{minipage}[t]{0.38\linewidth}
\centering
\centerline{\epsfig{figure=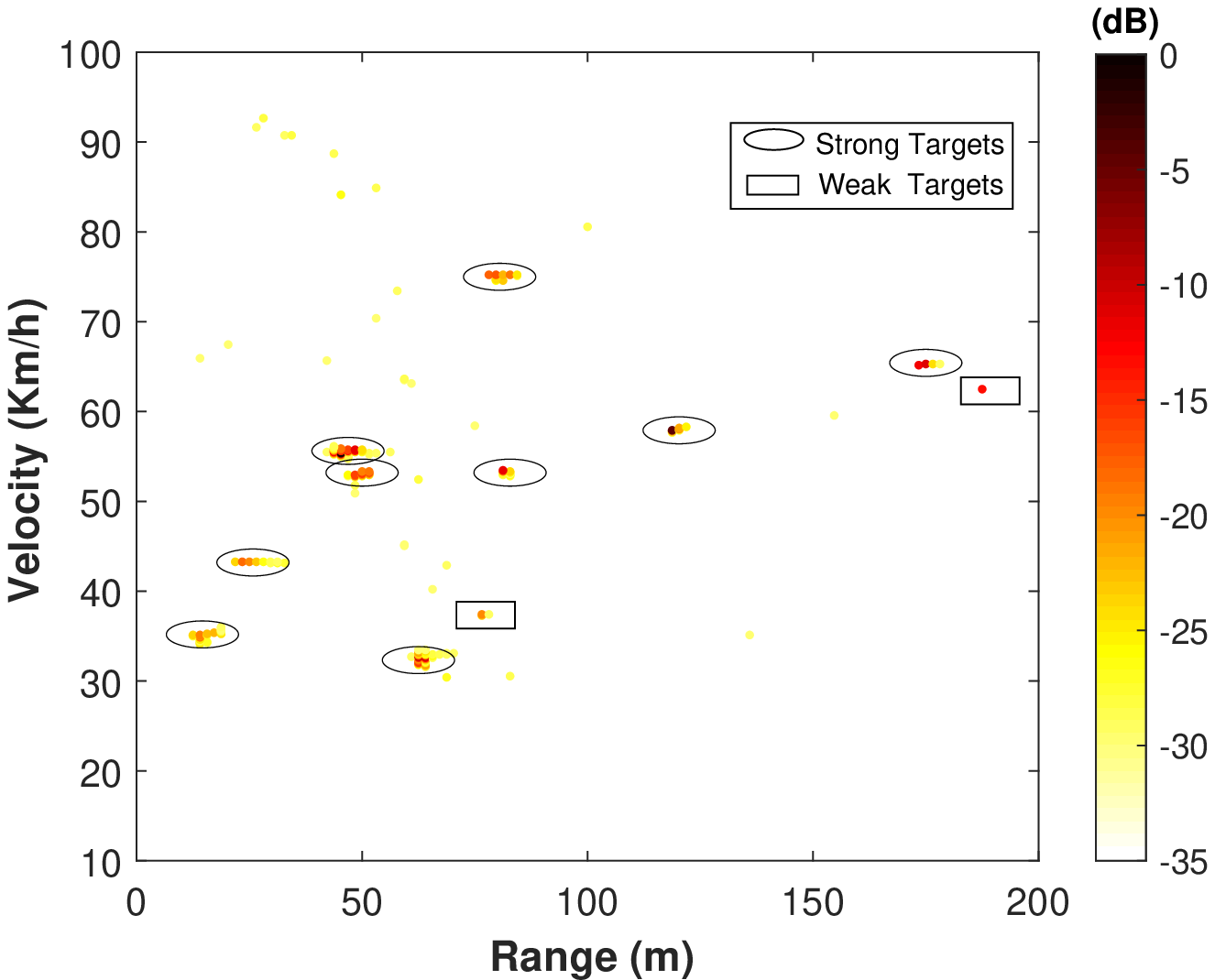,width=6.8cm}}
\end{minipage}}
\caption{Range-Doppler images of LFMCW radar, obtained applying (a) 2-D FFT and (b) the conventional IAA, to the original high-precision measured  data. }
\label{fig:unqlfmcw}
\end{figure}
\noindent In the 1bLIKES image, shown in Fig. \ref{fig:pmcw_likes}, the location and power of each target are accurately estimated.  In Fig. \ref{fig:pmcw_slim}, we show the range-Doppler image of 1bSLIM. 1bSLIM misses two weak targets basically because the 1bSLIM image is  too sparse.  Finally, 1bIAA,  as shown in  Fig. \ref{fig:pmcw_iaa}, provides accurate range-Doppler imaging results and identifies all  targets in the scene. Furthermore, 1bLIKES has noticeably more  relatively strong spurious peaks (clearly visible when viewed in color) than 1bIAA, making 1bIAA the best method for this example.
\begin{figure*}[htp]
\centering
\subfigure[1bPER]{
\label{fig:fmcw_1bperiodogram}
\begin{minipage}[t]{0.43\linewidth}
\centering
\centerline{\epsfig{figure=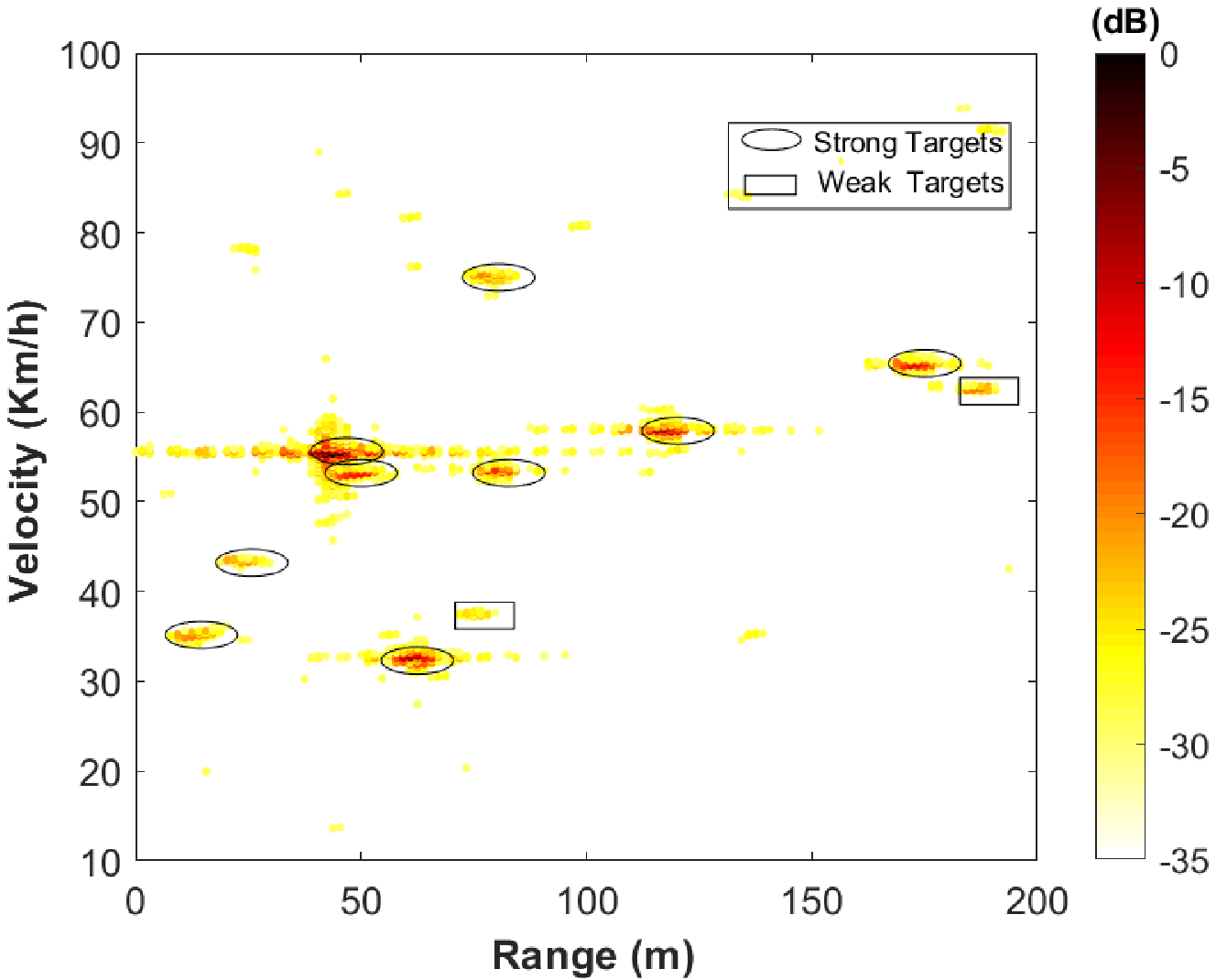,width=7.2cm}}
\end{minipage}}
\subfigure[1bSPICE]{
\label{fig:fmcw_1bspice}
\begin{minipage}[t]{0.43\linewidth}
\centering
\centerline{\epsfig{figure=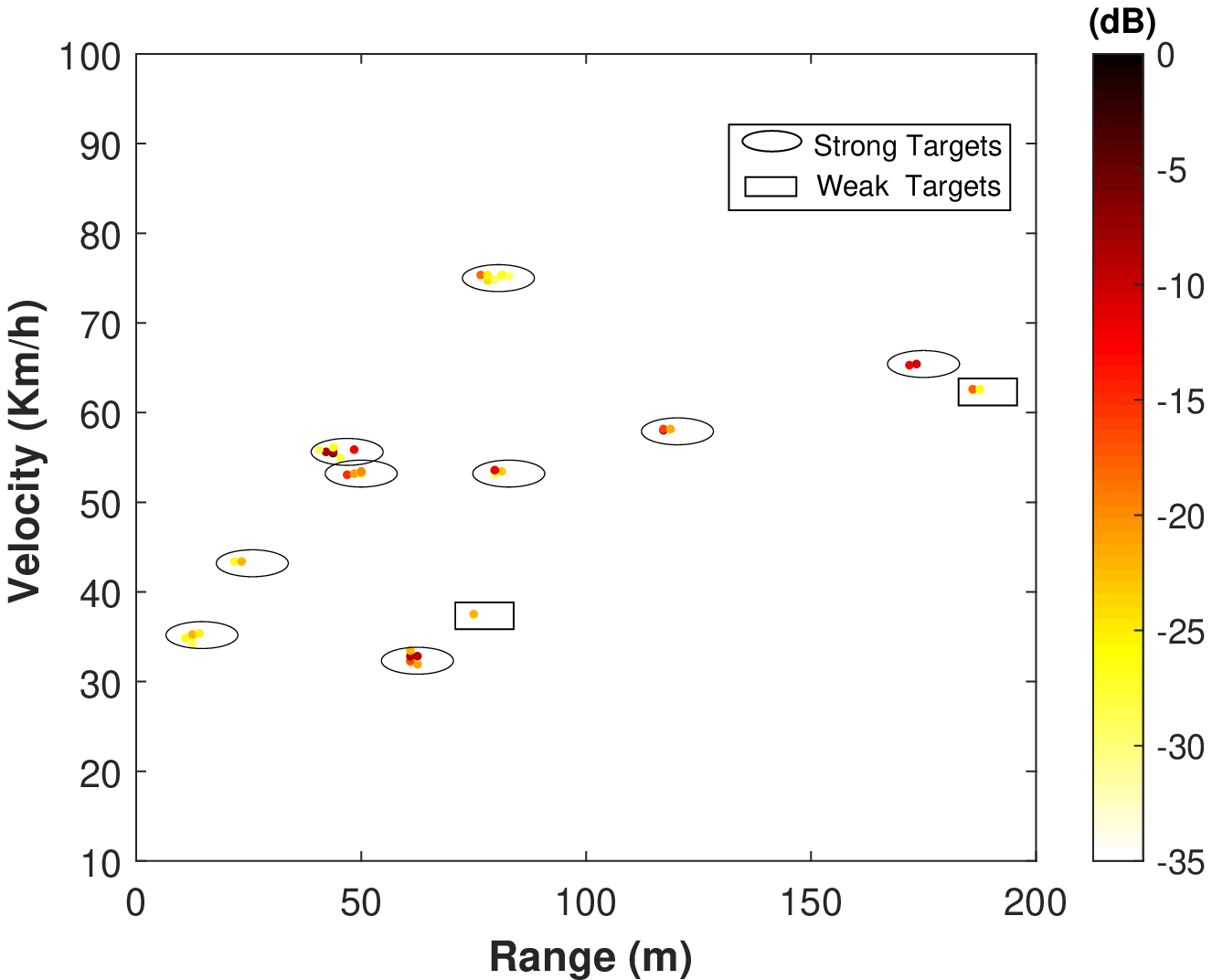,width=7.2cm}}
\end{minipage}}
\centering
\subfigure[1bLIKES]{
\label{fig:fmcw_likes}
\begin{minipage}[t]{0.43\linewidth}
\centering
\centerline{\epsfig{figure=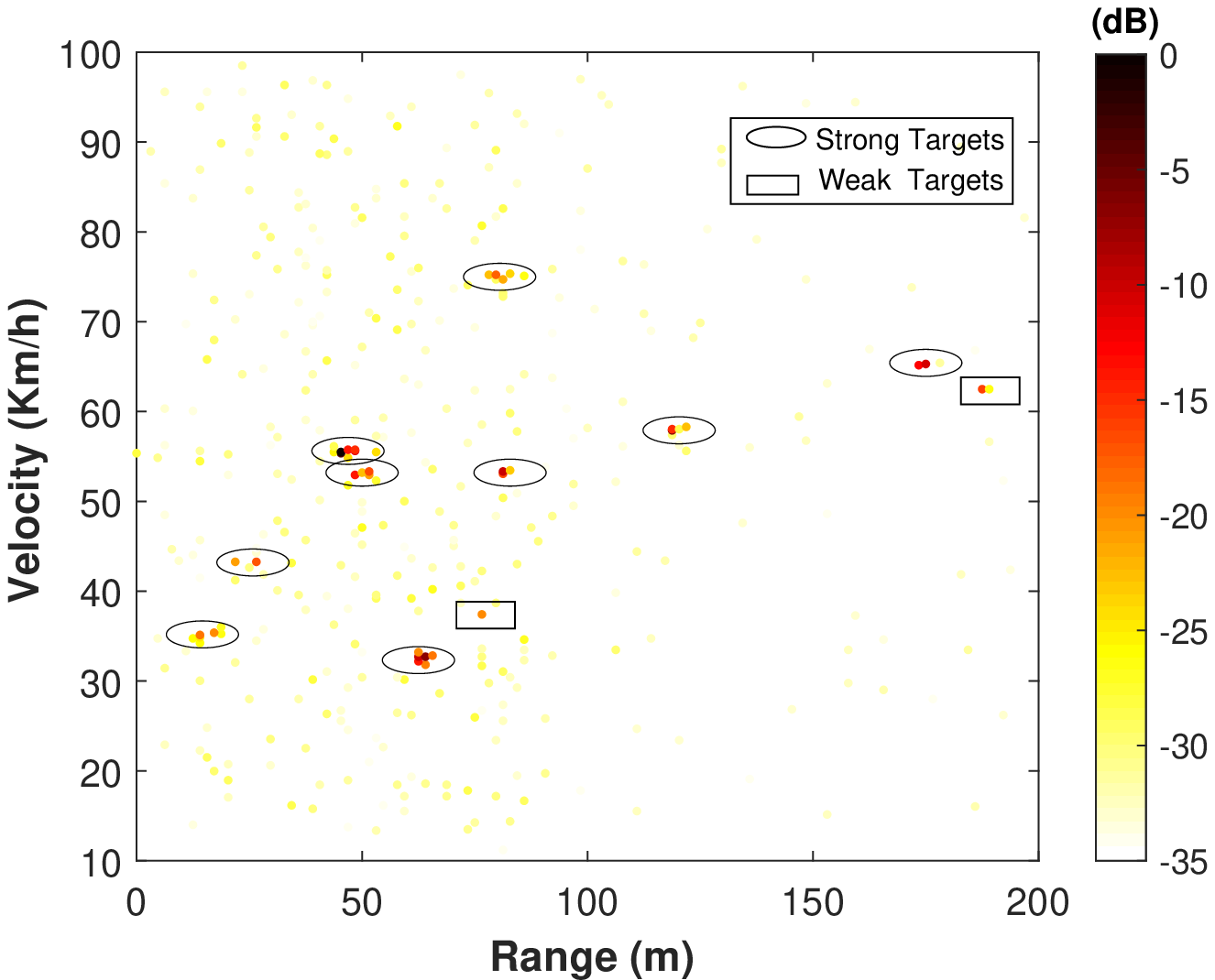,width=7.2cm}}
\end{minipage}}
\subfigure[1bSLIM]{
\label{fig:fmcw_slim}
\begin{minipage}[t]{0.43\linewidth}
\centering
\centerline{\epsfig{figure=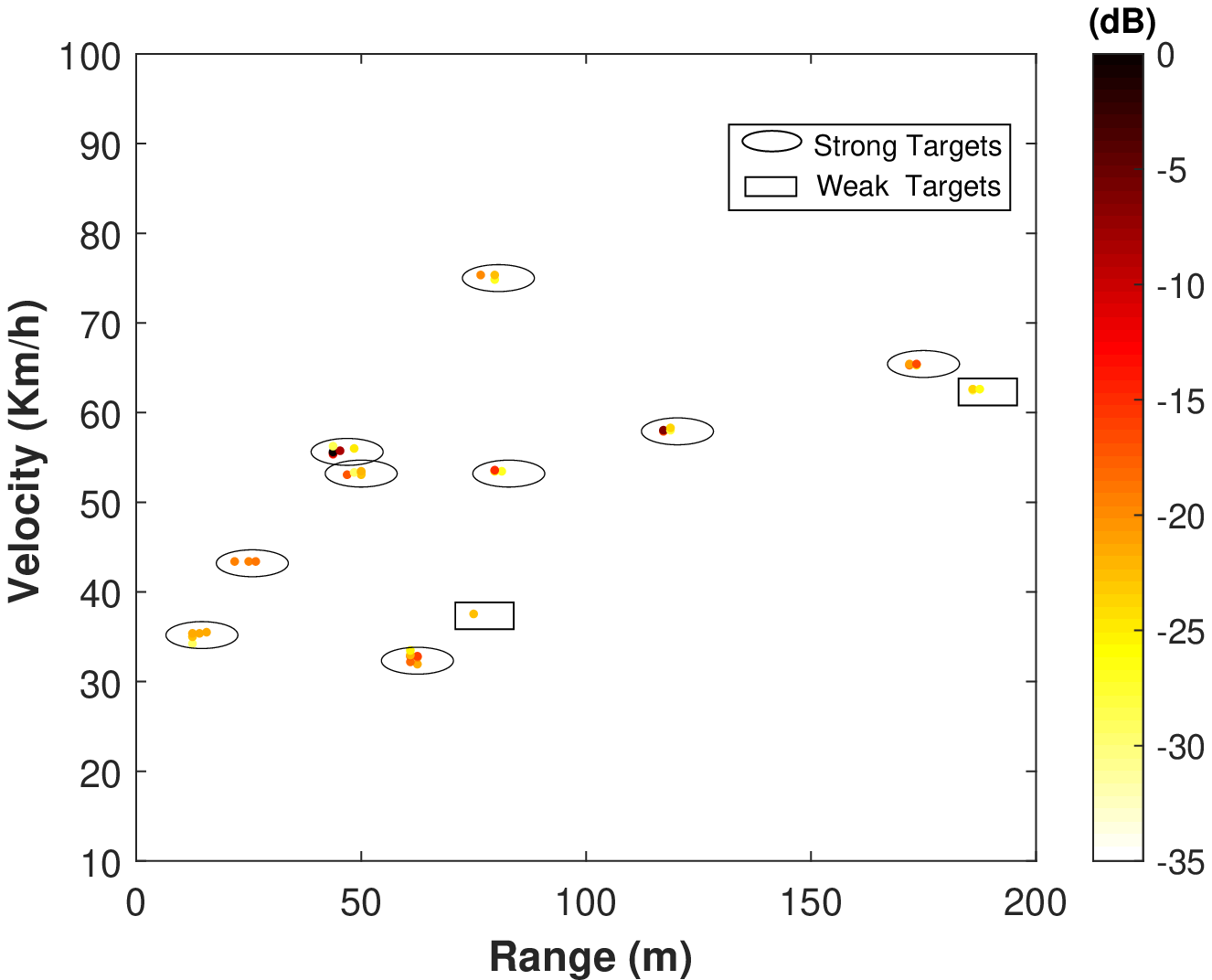,width=7.2cm}}
\end{minipage}}
\subfigure[1bIAA]{
\label{fig:fmcw_1biaa}
\begin{minipage}[t]{0.43\linewidth}
\centering
\centerline{\epsfig{figure=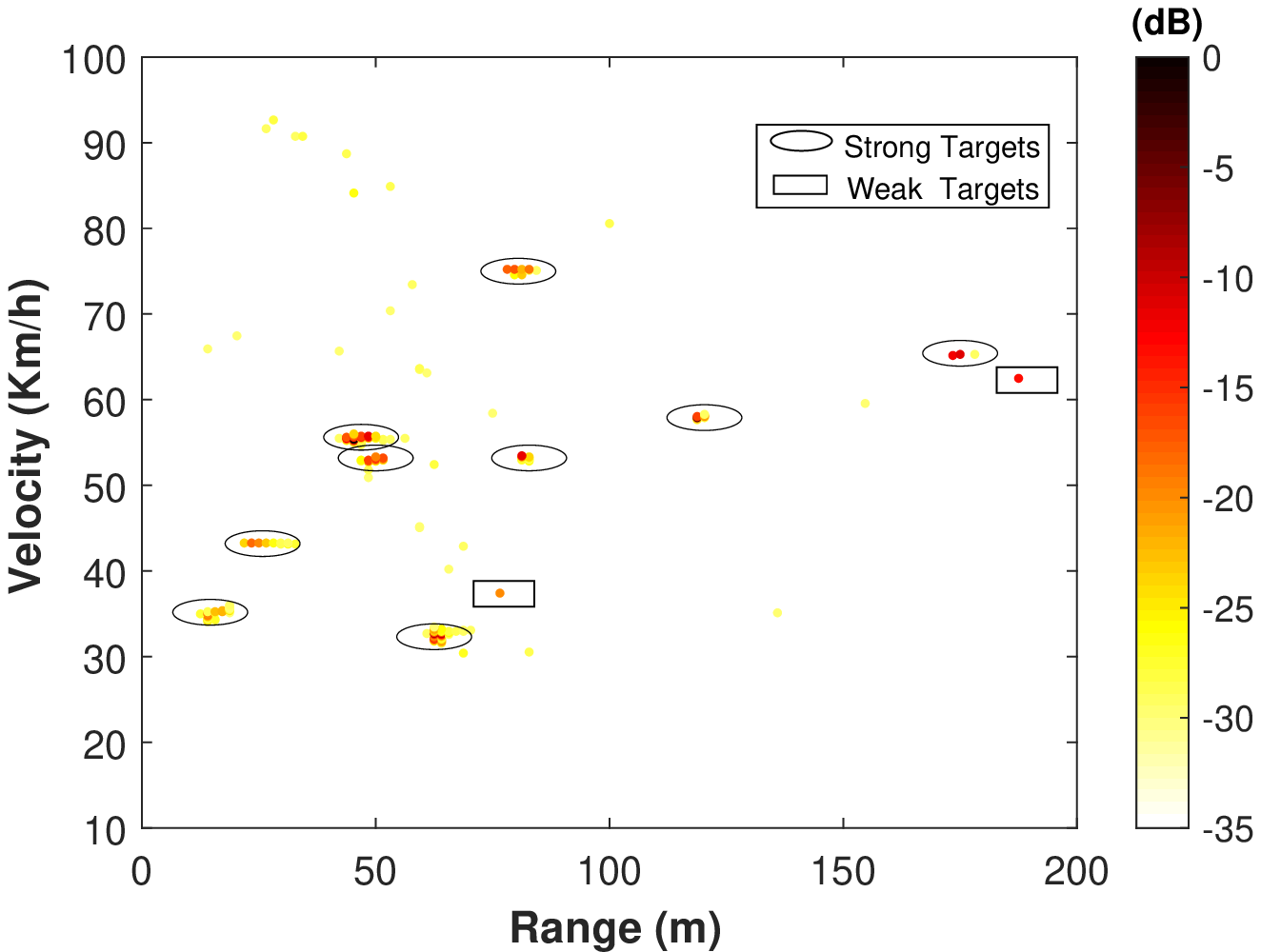,width=7.2cm}}
\end{minipage}}
\caption{Range-Doppler images of one-bit LFMCW radar,  obtained  applying (a) 1bPER, (b) 1bSPICE, (c) 1bLIKES, (d) 1bSLIM, (e) 1bIAA, to  the one-bit measured data.  }
\label{fig:1blfmcw}
\end{figure*}
\subsection{Experimental Example}
In this section, we  present an experimental example to demonstrate the performance of the proposed one-bit weighted SPICE algorithms for range-Doppler imaging via a one-bit LFMCW radar. The experimental data is collected using a 24 GHz radar sensor with bandwidth 25 MHz and PRI 80 ${\rm \mu s}$. The received signal is sampled by 16-bit ADCs and the  samples contain additive noise with unknown  variance. The dimensions of the 2-D data matrix are $N_1=64$ (for range) and $N_2=512$ (for Doppler). 
The number of grid points in the range and Doppler domains are set to $K_{\rm r} = 4N_1$, $K_{\rm d} = 4N_2$, respectively. Similar to the PMCW case in the previous example, we  use a PRI-varying threshold scheme that  keeps the threshold  constant within each chirp and only  changes the threshold from chirp to chirp.  Assume $A$ is the signal power in the high-precision data. Again, in practical systems, a rough estimate of the  received signal power can be obtained from an AGC circuit.   The one-bit data is obtained comparing the  high-precision data with a PRI-varying threshold whose real and imaginary parts are randomly selected from an eight-element set  $\lbrace -h_{\max}, -h_{\max}+\Delta, \dots, h_{\max}-\Delta, h_{\max} \rbrace$ with $h_{\max}=\frac{\sqrt{A}}{2}$ and $\Delta=\frac{2h_{\max}}{7}$.  \\
\indent In Figs. \ref{fig:fmcw_mf} and \ref{fig:fmcw_iaa}, we show the benchmark range-Doppler images  obtained by 2-D FFT and the conventional IAA applied to the high-precision data. Note that 9 strong targets and 2 weak targets are visible in both  2-D FFT and IAA images. In addition, a superposition of multiple point  scatterers is  identified by the radar as a single target. Compared with 2-D FFT, IAA drastically reduces the sidelobe levels and improves the resolution. For the present data size, the memory needed by the ADMM approach exceeds the maximum size limit of MATLAB, and hence  ADMM is not considered in this comparison.  Fig. \ref{fig:fmcw_1bperiodogram} shows  the range-Doppler image obtained by applying  1bPER  to the one-bit data.   Note that 1bPER smears the targets  significantly  and generates significant background clutter. Figs. \ref{fig:fmcw_1bspice}--\ref{fig:fmcw_1biaa} show the range-Doppler imaging results  obtained by the one-bit weighted SPICE algorithms applied to the one-bit data. All strong targets are accurately estimated and the  two weak targets are detected reasonably well by the proposed algorithms. Interestingly, the result of 1bIAA using one-bit data is quite similar to  that of the  conventional IAA using  high-precision data.  Furthermore, the proposed one-bit weighted SPICE algorithms applied to  one-bit data outperforms the 2-D FFT  approach applied to the high-precision data in terms of sidelobe levels and resolution, thereby demonstrating the effectiveness of the algorithms proposed in this paper. 
\section{Conclusions}
We have considered  range-Doppler imaging via  one-bit automotive LFMCW/PMCW radar systems, which use one-bit sampling with time-varying thresholds at the receiver to obtain signed measurements. We have formulated the  range-Doppler imaging  problem as  one-bit sparse-parameter estimation. We have first  presented a user parameter-free regularized minimization approach, namely 1bSLIM, to obtain accurate range-Doppler estimates. Inspired by the relationship between 1bSLIM and the original SLIM for high-precision data, we have also extended the conventional SPICE, LIKES and IAA to the  one-bit data case, referred  to as 1bSPICE, 1bLIKES and 1bIAA.  These four hyperparameter-free algorithms have been unified under the one-bit weighted SPICE umbrella. Computationally efficient implementations of the algorithms have been investigated as well.  Finally, in the simulated and experimental examples, we have compared the performance of the one-bit weighted SPICE algorithms to that of the one-bit periodogram and the ADMM log-norm approach and we  have shown that the proposed one-bit weighted SPICE methods are not only more accurate but  also  much faster.
\appendices
\section{Majorization-Minimization Approach} \label{appendix:A}
MM is a type of iterative technique, which can transform a difficult optimization problem into  a sequence  of simpler ones \cite{hunter2004tutorial,stoica2004cyclic,mairal2015incremental,hong2015unified,sun2016majorization}. Consider the following optimization problem:
\begin{equation}
   \mathop{\rm {minimize}}\limits_{\vec{x}} f(\vec{x}). \label{eq:objective}
\end{equation}
For a given feasible  initialization $\hat{\vec{x}}^0$, an MM algorithm for solving \eqref{eq:objective} is composed of two steps at the $t$th iteration: the \textbf{majorization} step and the \textbf{minimization} step.  The \textbf{majorization} step is to  find a function $g(\vec{x} | \hat{\vec{x}}^t)$ that  satisfies the following properties:
\begin{align}
    g(\hat{\vec{x}}^t| \hat{\vec{x}}^t)&=f(\hat{\vec{x}}^t), \\
    g(\vec{x}| \hat{\vec{x}}^t) &\geq f(\vec{x}),
\end{align}
where $\hat{\vec{x}}^t$ is the estimate of $\vec{x}$ at the $t$th iteration. The minimization of $g(\vec{x}| \hat{\vec{x}}^t)$, which is referred to as a majorizing function, should be easier than that of $f(\vec{x})$. The \textbf{minimization} step is to update  $\vec{x}$ via the minimization of $g(\vec{x}|\hat{\vec{x}}^t)$:
\begin{equation}
    \hat{\vec{x}}^{(t+1)}=\arg \mathop{\min}_{\vec{x}} g(\vec{x}|\hat{\vec{x}}^t).
\end{equation}
The  objective  function $f(\vec{x})$ is guaranteed  to  decrease monotonically  at each MM iteration because:
\begin{equation}
    f(\hat{\vec{x}}^t)=g(\hat{\vec{x}}^t| \hat{\vec{x}}^t) \geq g(\hat{\vec{x}}^{(t+1)}|\hat{\vec{x}}^t)\geq f(\hat{\vec{x}}^{(t+1)}).
\end{equation}
\section{Proof of the Inequality (\ref{ineq:lemma1})} \label{appendix:B}
Assume $\vec{P} \succ 0$.  Then, we have the inequality
\begin{small}
\begin{align}
\vec{D}& =    \left[ 
\begin{matrix}
  (\hat{\vec{R}}^t)^{-1}\Vec{A}\hat{\vec{P}}^t\Vec{P}^{-1}\hat{\vec{P}}^t\vec{A}^H(\hat{\vec{R}}^t)^{-1} & \vec{I} \\
  \vec{I}  &  \vec{R} 
\end{matrix} 
    \right] \nonumber \\
    &= 
    \left[ 
    \begin{matrix}
      (\hat{\vec{R}}^t)^{-1}\vec{A}\hat{\vec{P}}^t\vec{P}^{-\frac{1}{2}} \\
      \vec{A}\vec{P}^{\frac{1}{2}}
    \end{matrix} 
    \right]
    \left[
    \begin{matrix}
      \vec{P}^{-\frac{1}{2}}\hat{\vec{P}}^t\vec{A}^H(\hat{\vec{R}}^t)^{-1} & \vec{P}^{\frac{1}{2}}\vec{A}^H
    \end{matrix}
    \right]     \succeq 0.
\end{align}
\end{small}
By the properties of the Schur complement of $\vec{R}$ in $\vec{D}$, $\vec{D} \succeq 0 $ implies
\begin{equation}
    \Vec{R}^{-1}\preceq (\hat{\vec{R}}^t)^{-1}\Vec{A}\hat{\vec{P}}^t\Vec{P}^{-1}\hat{\vec{P}}^t\vec{A}^H(\hat{\vec{R}}^t)^{-1}.
\end{equation}
\bstctlcite{IEEEexample:BSTcontrol}
\bibliographystyle{IEEEtran} 
\normalsize
\bibliography{ref}
\end{document}